\UseRawInputEncoding
\documentclass[a4paper]{article}
\usepackage{amsmath, amsfonts,amssymb}
\usepackage[dvips]{graphicx}
\usepackage[T1]{fontenc}
\allowdisplaybreaks
\numberwithin{equation}{section}

\begin{document}
\thispagestyle{empty}
\noindent\emph{Marko V. Lubarda}\footnote{
M.V. Lubarda, Department of Mechanical and Aerospace Engineering, University of California, San Diego, USA.}\\
\emph{Vlado A. Lubarda}\footnote{V.A. Lubarda, NanoEngineering Department, University of California, San Diego, USA.}

\begin{center}
\Large\textsc{\bf On the motion of an evaporating respiratory droplet}
\end{center}

\centerline{\bf A b s t r a c t} \vskip3mm

An analysis of the projectile motion in stagnant air is presented for an evaporating respiratory micro-droplet which has been ejected from the mouth as an isolated droplet. It is assumed that the air resistance is a nonlinear function of the droplet's velocity and that the rate of decrease of the droplet's external surface area depends only on the relative humidity and the ambient temperature.  The droplet's initial content is considered to be 98 wt\% water, 1 wt\% salt and 1 wt\% protein. The change of the average density of the droplet due to water evaporation is determined, up to the instant when the droplet reduces to its nucleus, consisting of salt and dry protein only. The numerical solution of the governing differential equations of droplet's motion gives the trajectories of different-sized droplets ejected at different velocities and angles, and under different relative humidities and rates of evaporation.
The evaporation times are compared with the times for droplets to reach the ground after being ejected from a given height.
The maximum horizontal and vertical distances reached by the droplet are evaluated in the presence of wind and discussed in the context of possible infection spreading.


\section{Introduction}

There has been a large amount of research devoted to biophysical and chemical aspects of the investigation of infection spreading by inhalation of airborne droplets containing bacteria or viruses, which has recently intensified due to the pandemic caused by SARS-Cov-2 (Stadnytskyi et al., 2020; Balachandar et al., 2020; Jayaweera et al., 2020; Lieber et al., 2021).
Normal speaking and expiratory activities can emit thousands of respiratory droplets per second.
The volume and droplet count increase with loudness; loud speaking and coughing emit even more droplets.
According to early measurements by Duguid (1946),
95\%  of the emitted respiratory droplets had the diameter between
2 and 100 $\mu$m, the most common diameter being between 4 and 8 $\mu$m.
Similar size distributions were found in droplets produced by sneezing, coughing, and normal speaking; the smaller droplets were relatively
more numerous in the case of sneezing.
More recent measurements with sub-micron resolution indicate that 80--90\% of particles produced by human expiratory activities are
smaller than 1 $\mu$m (Papineni and Rosenthal, 1997). The measured size, however, could correspond
to the size of the dry droplet residue (Morawska et al., 2009), because the time
spent in the air before the droplet's detection and its size measurement could have been long enough for significant evaporation to take place;
see also Xie et al. (2009) and Tellier et al. (2019). The fractional viral load of respiratory droplets, i.e., the
average number of virions per droplet, has been discussed by Mittal et al. (2020). The motion of respiratory droplets can also be affected by their interaction with other particles of non-biological origin present in the air (Morawska et al., 2006).

Small droplets (say less than 10 $\mu$m initial diameter, further decreasing in size due to evaporation) are easily carried by a puff of exhaled air
produced by breathing, talking, coughing, or sneezing, with small relative velocity, and can remain afloat for long time; they can be further convected by outside wind or indoor draft or ventilation (Liu and Novoselac, 2014, Li et al., 2018; Wang
et al., 2019; Anchordoqui and Chudnovsky, 2020; Cheng et al., 2020). As a result, the probability of their inhalation and thus infection, if they contain viruses such as SARS-CoV-2, substantially increases.  The Brownian motion of very small (aerosol) droplets has also been studied using the Langevin differential
equation with a stochastic term due to random interactions of the droplet with the individual molecules of
the air (Das et al., 2020). On the other hand, large droplets (say, greater than 50 or 100 $\mu$m-diameter),
if ejected from the mouth into the air outside of the exhaled puff, move like projectiles. These droplets are rapidly slowed down by drag resistance from the ambient air, and fall to the ground or nearby surfaces or people, which
presents a risk of infection when such droplets are touched and transmitted to mouth or eyes. If large droplets are  ejected within the exhaled jet stream, they initially move within the advancing and slowing puff, but can quickly exit from it due to their large weight and developed velocity relative to the puff (Xie et al., 2007; Wang et al., 2020). Once outside the puff,
these droplets move as isolated projectiles with initial positions and speeds determined from the more involved analysis describing their motion within the moving and expanding puff, and corresponding to the instant when they  exit the puff. Larger droplets may also be potentially more infectious due to higher viral
content (Li et al., 2021). The trajectories of intermediate-size droplets are more complex as they move within the puff longer distances and may evaporate to their nucleus size while still within the puff (Balachandar et al., 2020; Bourouiba, 2020; Giovanni, 2020; Vuorinen et al., 2020).

In this paper, we present an analysis of the projectile motion of an isolated micro-droplet ejected from the mouth at different angles and initial velocities. The air resistance is
represented by a nonlinear function of the droplet's velocity, and the rate of decrease of the droplet's external surface area is assumed to depend only on the relative humidity
and the ambient temperature. The droplet's initial content is considered to be 98 wt\% water, 1 wt\% salt (NaCl), and 1 wt\% protein (mucus). The change of the average density of the droplet caused by water evaporation is determined, up to the instant when all water evaporates and the droplet reduces to its nucleus (droplet residue), consisting of salt and dry protein only. This also specifies the change of the droplet's mass, which enables numerical solution of the governing differential equations of the droplet's motion, for an arbitrary angle of ejection and initial velocity, corresponding to either soft or loud speaking, coughing, or sneezing. The trajectories of different-sized droplets are determined under different relative humidities.
In each case the maximum horizontal distance traveled by the droplet in stagnant air is rapidly reached, which is followed by essentially vertical descent of the droplet. The maximum horizontal reach in soft talking is compared to that in coughing. We also evaluate and compare the time for the droplet to evaporate down to its nucleus size and the time for the droplet to reach the ground when ejected from a given height. The effects of the angle of ejection of the droplet on its maximum horizontal and vertical reach and the entire trajectory are discussed. The wind effects are also included in the analysis and discussed in the context of infection spreading by transmission of viruses and other pathogens via respiratory droplets.

\section{Projectile motion of an evaporating droplet with a nonlinear drag force}

For smaller droplets and for sufficiently small initial velocities, the Reynolds number of an isolated droplet is bellow 1 and the linear Stokes' type drag model can be adopted in the study of such droplet's motion. For larger droplets (say, greater than about 5 microns) and for larger initial velocities (e.g., velocities about 10 m/s as during coughing, or up to 50 m/s as during sneezing), the Reynolds number, at least in the early stage of the droplet's motion, is in the range 1<Re<1000, and the drag force becomes a nonlinear function of velocity. To encompass the entire range of Reynolds number, the drag force in still air is commonly written as
\begin{equation}\label{2.1}
{\bf F}_{\rm d}=-\frac{1}{2}\,\rho_{\rm air} Ac_{\rm d}v\boldsymbol{v}\,,\quad A=\pi R^2\,,
\end{equation}
where the drag coefficient $c_{\rm d}$ depends on the Reynolds number, in accord to experimental data (Schlichting, 1979; White, 2006). The air density is $\rho_{\rm air}$, the current radius of the droplet is $R$, its velocity vector is $\boldsymbol{v}$, and $v$ is its magnitude. The following expression fits this data sufficiently well in the
range $0.2<{\rm Re}\le 3\times 10^5$ (Lapple, 1951; Khan and Richardson, 1987)
\begin{equation}\label{2.2}
c_{\rm d}=
\begin{cases}
24{\rm Re}^{-1}+3{\rm Re}^{-0.28}, & 0.2\le{\rm Re}\le 10^3\,,\\
0.458\,,& 10^3\le {\rm Re}\le 3\times 10^5\,.
\end{cases}
\end{equation}
Below ${\rm Re}=0.2$, the dominant contribution to $c_{\rm d}$ is $24/{\rm Re}$, which gives rise to linear drag force, proportional to velocity (Stokes' flow).
Other fitting expressions have also been proposed and used, e.g., Zhu et al. (2006), Xie et al. (2007), Wang et al. (2020), Cheng et al. (2020), and Lieber et al. (2021). For example,
Xie et al. (2007), used $c_{\rm d}=24{\rm Re}^{-1}+4{\rm Re}^{-1/3}$ for ${\rm Re}\le 1000$ and $c_{\rm d}=0.424$ for ${\rm Re}\ge 1000$.
The Reynolds number at an arbitrary time $t$ during the motion of evaporating droplet is defined by
\begin{equation}\label{2.3}
{\rm Re}(t)=\frac{2R(t)v(t)}{\nu_{\rm air}}\,.
\end{equation}
The kinematic viscosity of air is $\nu_{\rm air}=\eta_{\rm air}/\rho_{\rm air}$, where $\eta_{\rm air}$ is its dynamic viscosity. Typically, the Reynolds number of a relatively large droplet is below 1,000; for example, for $\nu_{\rm air}=1.516\times 10^{-5}\,{\rm m}^2/s$ and at the instant when $v=10$ m/s and $R=100\,\mu{\rm m}$, the Reynolds number is ${\rm Re}\approx 132$. On the other hand, for a small droplet of radius $R=5\,\mu{\rm m}$ and current velocity $v=1$ m/s, the Reynolds number is ${\rm Re}\approx 0.66$. If the air surrounding a droplet moves with the velocity $\boldsymbol{v}_{\rm a}$, the relative velocity $\boldsymbol{v}-\boldsymbol{v}_{\rm a}$ should be used in place of the total velocity in the expression for the drag force (\ref{2.1}) and the definition of the Reynolds number (\ref{2.3}).

\begin{figure}
\centering{
\includegraphics[scale=0.75]{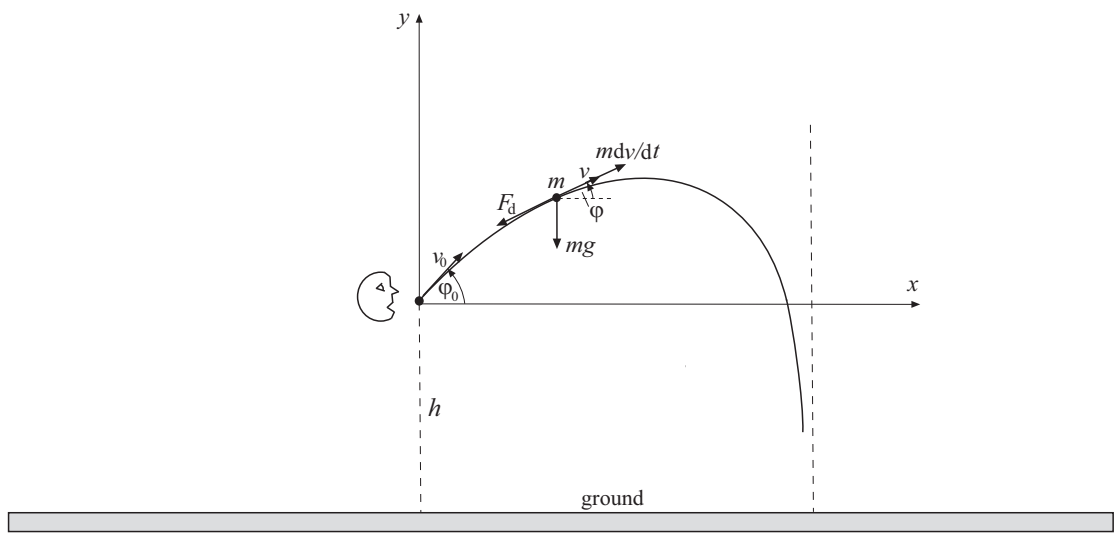}
}
\caption{\label{Fig1} The motion of the droplet of initial mass $m_0$ ejected at the height $h$ above the ground with the initial velocity $v_0$ at an angle $\varphi_0$ with respect to the horizontal $x$ axis. In the position shown at time $t$,
the forces acting on the droplet are its weight $mg$ and the drag force $F_{\rm d}$, opposite to its current velocity $v$. The horizontal and vertical components of velocity are $v_x=v\cos\varphi$ and $v_y=v\sin\varphi$.}
\end{figure}

Suppose that the droplet having the initial mass $m_0$ and corresponding initial radius $R_0$ is emitted from a mouth at the height $h$
above the ground with the initial velocity $v_0$ at the angle $\varphi_0$ relative to the horizontal $x$ direction (Fig. \ref{Fig1}). At an arbitrary time $t$ during its projectile-type motion, the mass of the droplet is $m<m_0$, its radius $R<R_0$, and its velocity $v<v_0$. The current inclination angle of the velocity vector is $\varphi$.
The vectorial form of the differential equation of the droplet's motion in still air is then
\begin{equation}\label{2.4}
m\,\frac{\mathrm{d}{\bf v}}{\mathrm{d}t}={\bf F}_{\rm d}+m{\bf g}-{\bf u}_{\rm rel}\,\frac{\mathrm{d}m}{\mathrm{d}t}\,.
\end{equation}
The average ejection velocity, relative to the droplet, of all particles leaving the droplet due to evaporation is ${\bf u}_{\rm rel}={\bf u}-{\bf v}$,
and ${\bf g}=\{0,-g\}$ is the gravitational acceleration. The buoyancy force is ignored as the density of a respiratory droplet is about 1000 times greater than the density of the air.
By assuming that ${\bf u}_{\rm rel}=0$, the governing differential equations for the velocity components $(v_x,v_y)$ of the droplet are
\begin{equation}\label{2.5}
m\,\frac{\mathrm{d}v_x}{\mathrm{d}t}=-F_{\rm d}\,\frac{v_x}{v}\,,\quad
m\,\frac{\mathrm{d}v_y}{\mathrm{d}t}=-F_{\rm d}\,\frac{v_y}{v}-mg\,,
\end{equation}
where $F_{\rm d}=\rho_{\rm air} Ac_{\rm d}v^2/2$ from (\ref{2.1}).
Upon the substitution of (\ref{2.1}) and the division by the current mass $m=\rho V$, where $\rho$ is the (average) mass density of the droplet at time $t$ and $V=4\pi R^3/3$ is the
droplet's current volume, (\ref{2.5}) becomes
\begin{equation}\label{2.6}
\frac{\mathrm{d}v_x}{\mathrm{d}t}=-\frac{3\rho_{\rm air}}{8\rho}\,\frac{c_{\rm d}}{R}\,v_xv\,,\quad
\frac{\mathrm{d}v_y}{\mathrm{d}t}=-\frac{3\rho_{\rm air}}{8\rho}\,\frac{c_{\rm d}}{R}\,v_yv-g\,,
\end{equation}
with the drag coefficient $c_{\rm d}$ defined by (\ref{2.2}). The accompanying initial conditions are $v_x(0)=v_0\cos\varphi_0$ and $v_y(0)=v_0\sin\varphi_0$.

Equations (\ref{2.6}) are two coupled differential equations, which require numerical integration, provided that the expressions for $R=R(t)$ and $\rho=\rho(t)$ are determined,
corresponding to the rate by which the mass of the droplet evaporates, as elaborated upon in section 3.
After the velocity components $v_x$ and $v_y$ have been determined, the $(x,y)$ coordinates of the trajectory of the droplet follow by numerical quadrature from $x(t)=\int_0^tv_x(\theta)\,\mathrm{d}\theta$ and  $y(t)=\int_0^tv_y(\theta)\,\mathrm{d}\theta$, as reported in section 4.

\subsection{Motion of the droplet's residue}

At some instant of motion ($t_*$), all water from a respiratory droplet evaporates and a droplet is reduced to its nucleus (salt and dry protein residue) of the radius denoted by $R_*$. From that instant on, i.e., for time $t\ge t_*$, the motion of this non-volatile residue proceeds as the motion of a non-evaporating spherical particle of constant radius $R_*$. The initial conditions for this
motion are $v_x(t=t_*)=v_*\cos\varphi_*$, $v_y(t=t_*)=v_*\sin\varphi_*$, and $x(t=t_*)=x_*$, $y(t=t_*)=y_*$. The quantities $(t_*,v_*,\varphi_*,x_*,y_*)$ have been
determined from the preceding analysis of the motion of the evaporating droplet, and correspond to the end of that part of motion, i.e., the instant when the droplet evaporates to its nucleus size of radius $R_*$. The numerical analysis of the motion of evaporating droplet shows that the Reynolds number rapidly diminishes with time, after the droplet's emission from the mouth, because the
droplet's velocity rapidly diminishes, as well as its radius. As a consequence, the motion of the droplet's residue is in many cases the motion under linear (Stokes')
drag force, $F_{\rm d}=cv$, where $c=6\pi\eta_{\rm air}R_*$. It then
readily follows that the velocity components of the droplet's residue are
\begin{equation}\label{2.7}
\begin{split}
v_x(t)&=v_*\cos\varphi_*\,e^{-k_*(t-t_*)}\,,\quad k_*=\frac{c}{m_*}\,,\quad t\ge t_*\,,\\
v_y(t)&=\left(v_{\rm t}+v_*\sin\varphi_*\right)e^{-k_*(t-t_*)}-v_{\rm t}\,,\quad v_{\rm t}=\frac{g}{k_*}\,,
\end{split}
\end{equation}
where $m_*=\rho_* V_*$ is the mass of the droplet's residue and $V_*=4\pi R_*^3/3$ is its volume. The velocity $v_{\rm t}$ is the so-called terminal velocity, at which the drag force balances the gravity force ($cv_{\rm t}=m_*g$). The corresponding coordinates of the droplet's residue are given by
\begin{equation}\label{2.8}
\begin{split}
x(t)-x_*&=\frac{1}{k_*}\,v_*\cos\varphi_*\left[1-e^{-k_*(t-t_*)}\right],\quad t\ge t_*\,,\\
y(t)-y_*&=\frac{1}{k_*}\left(v_{\rm t}+v_*\sin\varphi_*\right)\left[1-e^{-k_*(t-t_*)}\right]-v_{\rm t}(t-t_*).
\end{split}
\end{equation}
By eliminating  $t-t_*$ in (\ref{2.8}), the $y=y(x)$ representation of the residue trajectory can also be explicitly given.
For sufficiently large times ($t\gg t_*$), the residue approaches the vertical asymptote $x=x_*+(v_*/k_*)\cos\varphi_*$, with the residue's time-dependent position $y$ being given by
\begin{equation}\label{2.9}
y(t)=y_*+\frac{1}{k_*}\,(v_{\rm t}+v_*\sin\varphi_*)-v_{\rm t}(t-t_*)\,.
\end{equation}

\section{Evaporation of a droplet}

Suppose that a respiratory droplet is exhaled from mouth as a spherical particle of initial radius $R_0$ and the corresponding mass $m_0$. It will be assumed that $98\%$ of $m_0$ is pure water ($m_{\rm w}^0=0.98m_0$), $1\%$ a salt (NaCl), and $1\%$ a protein (mucin), i.e.,
\begin{equation}\label{3.1}
m_0=m_{\rm w}^0+m_{\rm s}+m_{\rm p}=m_{\rm w}^0+f_{\rm s}m_0+f_{\rm p}m_0\,,\quad (f_{\rm s}=0.01\,,\,f_{\rm p}=0.01)\,.
\end{equation}
This initial content of the respiratory saliva droplet is taken somewhat arbitrarily , from a large variety of data found in the literature
(Sarkar et al. 2019; Chaudhuri et al., 2020; Netz, 2020; Stadnytskyi, 2020; Lieber, 2021), but the analysis proceeds similarly if other initial droplet's content is assumed, with the main effect on the size of the droplet's residue.
Furthermore, we assume that the salt and protein are non-volatile at common ambient temperatures ($m_{\rm s}={\rm const.}$ and $m_{\rm p}={\rm const.}$), and that only water evaporates during the droplet's motion, i.e., the time rate of the droplet's mass is equal to the time rate of water content in the droplet ($\mathrm{d}m/\mathrm{d}t=\mathrm{d}m_{\rm w}/\mathrm{d}t$). Initially, the water and salt in an exhaled droplet form an aqueous solution in which ${\rm Na}^+$ and ${\rm Cl}^-$ ions are dispersed throughout the water, with the initial mass (weight) concentration of salt
\begin{equation}\label{3.2}
\frac{m_{\rm s}}{m_{\rm sol}^0}=\frac{m_{\rm s}}{m_0-m_{\rm p}}=\frac{f_{\rm s}}{1-f_{\rm p}}\,.
\end{equation}
The objective is to determine the time-dependent decrease of the mass and density of the droplet during its evaporation.

The evaporation of a droplet is a complicated physical and chemical process that can be analyzed with various degrees of complexity and sophistication. This may
include the consideration of the energy equation and
the heat and mass transfer between the droplet and its surroundings, and the effects of the ambient temperature and the temperature of the droplet on its evaporation rate
(e.g., Kukkonen et al, 1989; Sun and Ji, 2007; Liu et al., 2019; Chaudhuri et al., 2020; Chen, 2020; Netz; 2020). In our present analysis, we adopt a simplified phenomenological approach in which it is assumed that during evaporation the time rate of the external surface of the droplet, $S(t)=4\pi R^2(t)$, is a given constant ($s$), dependent only on the relative humidity (R.H.) and the ambient temperature ($T_{\rm amb}$). This is so because the evaporation is driven by the difference between the concentration of the water vapor at the surface of the droplet and in the surrounding air. Thus, we assume that
\begin{equation}\label{3.3}
\frac{\mathrm{d}S}{\mathrm{d}t}=-s\,,\quad s=({\rm R.H.},T_{\rm amb})\,.
\end{equation}
A possible effect of the velocity of the moving droplet on its evaporation rate $s$ is not included in (\ref{3.3}).
Surface evolution expression of the type (\ref{3.3}) has also been adopted in the early work by Wells (1934), and in the later work by Kayser and Bennett (1977). Alternative expressions could also be adopted, such as $\mathrm{d}m/\mathrm{d}t=-\kappa$ and $\mathrm{d}m/\mathrm{d}t=-\mu S$, where $\kappa$ and $\mu$ depend on the ambient and droplet's temperatures and other
physical parameters of the evaporation process. The former expression was derived by Kukkonen et al. (1989) and later used by Cheng et al. (2020).

Returning to the rate expression (\ref{3.3}), to determine the expression $s=s({\rm R.H.})$, we use the results from the analysis of evaporation kinetics of water droplets by Su et al. (2018). From their Fig. 5, the  initial radius of a spherical droplet $R_0=25\,\mu$m is reduced by evaporation to the radius $R=5\,\mu$m in approximately 2.2 seconds for R.H.=0\%, 2.75 seconds for R.H.=20\%,  4.5 seconds for R.H.=50\%, 11.5 seconds for R.H.=80\%, and in about 23 seconds for R.H.=90\%. The ambient temperature was $T_{\rm amb}=293$ K. By using this, from (\ref{3.3}) it follows that the corresponding values of $s$ are $\{3.427, 2.742, 1.676, 0.656, 0.328\}\times 10^{-9}\,{\rm m^2/s}$. These values are well-interpolated by a linear fit $s=s_0(1-{\rm R.H.}/100)$, where  $s_0=3.427\times 10^{-9}\,{\rm m^2/s}$.
Since saliva droplets evaporate little slower than pure water droplets, the value of $s_0$ can be taken a little lower, and we adopt the following expression
\begin{equation}\label{3.4}
s=s_0(1-{\rm R.H.}/100)\,,\quad s_0=3.25\times 10^{-9}\, {\rm m^2/s}\,.
\end{equation}
For other than room ambient temperatures, the expression (\ref{3.4}) may still approximately apply by adjusting the value of the temperature-dependent coefficient $s_0=s_0(T_{\rm amb})$.
Because the evaporation is more rapid at higher ambient temperature, the coefficient $s_0$ is expected to be a monotonically increasing function of $T_{\rm amb}$.

By integrating (\ref{3.3}), it follows that
\begin{equation}\label{3.5}
S(t)=S_0\left(1-\frac{t}{\tau}\right),\quad  R(t)=R_0\left(1-\frac{t}{\tau}\right)^{1/2},\quad \tau=\frac{S_0}{s}\,,\quad S_0=4\pi R_0^2\,.
\end{equation}
These expressions apply until all water evaporates and droplet reduces its radius to the radius of its droplet nucleus. The square-root dependence of $R$ on time $t$ in (\ref{3.5}) can be compared with the well-know `radius-square-law' relation between initial and current radius of the droplet $R^2-R_0^2=kt$ ($k={\rm const.}$), e.g., Jakubczyk (2012), Netz (2020), and Balachandar et al. (2020).

The salt is soluble in water up to its weight concentration of 0.357. In this range of concentration the experimental data on the density of aqueous solution of NaCl is well reproduced by a linear fit
\begin{equation}\label{3.6}
\rho_{\rm sol}^{\rm aq}=\rho_{\rm w}+\kappa\,\frac{m_{\rm s}}{m_{\rm sol}^{\rm aq}}\,,\quad 0\le \frac{m_{\rm s}}{m_{\rm sol}^{\rm aq}}\le 0.357\,,
\end{equation}
where $\rho_{\rm w}=997\,{\rm kg/m^3}$ is the mass density of water at room temperature, and $\kappa=755\,{\rm kg/m^3}$ is the slope of this linear fit. As the water evaporates from a droplet, the salt concentration increases, and after it reaches the saturation concentration of 0.357, the water and salt make a two-phase solution consisting of  a solid (crystalline) salt phase sedimented in the aqueous solution with salt concentration of 0.357. Finally, after all water evaporates, only salt remains, which, together with a dry protein, constitutes a droplet residue of mass
\begin{equation}\label{3.7}
m_*=m_{\rm s}+m_{\rm p}=(f_{\rm s}+f_{\rm p})m_0\,.
\end{equation}

\subsection{Density of the evaporating droplet}

As the respiratory droplet looses water by evaporation, its density increases. This affects the motion, as seen from eqs. (\ref{2.6}) in which the density $\rho$ appears in the denominator of both equations. We therefore proceed to determine the density of the droplet as a function of time. The initial radius of the droplet $R_0$ and, thus, its
initial volume $V_0=4\pi R_0^3/3$, are assumed to be known. If the rate of evaporation is specified by (\ref{3.5}), the volume of the droplet at
an arbitrary instant of evaporation is
\begin{equation}\label{3.8}
V(t)=\frac{4\pi}{3}\,R^3(t)\,,\quad R(t)=R_0\left(1-\frac{t}{\tau}\right)^{1/2},\quad \tau=\frac{4\pi R_0^2}{s}\,.
\end{equation}
The density expression will be first derived for the stage of the evaporation process during which the concentration of dissolved salt in water is below the saturation value of 0.357. At an arbitrary instant during this stage of evaporation, the mass of the droplet is
\begin{equation}\label{3.9}
m=m_{\rm sol}^{\rm aq}+m_{\rm p}=m_{\rm sol}^{\rm aq}+f_{\rm p}m_0\,,\quad m_{\rm sol}^{\rm aq}=m_{\rm w}+m_{\rm s}=m_{\rm w}+f_{\rm s}m_0\,.
\end{equation}
Furthermore, the volume of the droplet is taken to be the sum of the volumes of the protein and the aqueous salt solution, i.e.,
\begin{equation}\label{3.10}
V=V_{\rm sol}^{\rm aq}+V_{\rm p}=V_{\rm sol}^{\rm aq}+\frac{m_{\rm p}}{\rho_{\rm p}}\,,\quad \rho_{\rm p}=1,350\,{\rm kg/m^3}\,.
\end{equation}
The mass density of the dry protein is $\rho_{\rm p}=1,350\,{\rm kg/m^3}$, as frequently adopted in the literature (e.g., Fisher et al., 2004), and is considered to be constant. Thus, from (\ref{3.10}),
\begin{equation}\label{3.11}
V_{\rm sol}^{\rm aq}=V-\frac{f_{\rm p}}{\rho_{\rm p}}\,m_0\,.
\end{equation}

The initial mass of the droplet $m_0$ is still unknown, because we assumed that only the initial droplet's volume $V_0$ is given (as, for example, obtained by the measurement of the initial droplet's radius), although one may reasonably expect that $m_0$ is nearly equal to $\rho_{\rm w}V_0$, because of the small initial weight percents of salt and protein. By introducing the mass density of the aqueous salt solution ($\rho_{\rm sol}^{\rm aq}$), we can write
\begin{equation}\label{3.12}
m_{\rm sol}^{\rm aq}=\rho_{\rm sol}^{\rm aq}V_{\rm sol}^{\rm aq}\,,
\end{equation}
and the substitution of (\ref{3.6}) and (\ref{3.11}) into (\ref{3.12}) gives the quadratic equation for $m_{\rm sol}^{\rm aq}$,
\begin{equation}\label{3.13}
(m_{\rm sol}^{\rm aq})^2-b\,m_{\rm sol}^{\rm aq}-c=0\quad \Rightarrow\quad m_{\rm sol}^{\rm aq}=\frac{1}{2}\left(b+\sqrt{b^2+4c}\right),
\end{equation}
where
\begin{equation}\label{3.14}
b=\rho_{\rm w}\left(V-f_{\rm p}\,\frac{\rho_0}{\rho_{\rm p}}\,V_0\right),\quad c=\kappa f_{\rm s}\rho_0V_0\left(V-f_{\rm p}\,\frac{\rho_0}{\rho_{\rm p}}\,V_0\right).
\end{equation}

To determine the initial density $\rho_0$, we apply (\ref{3.13}) to the initial instant when  $(m_{\rm sol}^{\rm aq})^0=m_0-m_{\rm p}=(1-f_{\rm p})m_0=(1-f_{\rm p})\rho_0V_0$. Upon solving for $\rho_0$, it follows that
\begin{equation}\label{3.15}
\rho_0=\frac{(1-f_{\rm p})\rho_{\rm w}+\kappa f_{\rm s}}{
(1-f_{\rm p})^2+f_{\rm p}(1-f_{\rm p})\rho_{\rm w}/\rho_{\rm p}+\kappa f_{\rm s}f_{\rm p}/\rho_{\rm p}}\,.
\end{equation}
By substituting $f_{\rm s}=f_{\rm p}=0.01$, $\rho_{\rm w}=997\,{\rm kg/m^3}$, $\rho_{\rm p}=1,350\,{\rm kg/m^3}$, and $\kappa=775\,{\rm kg/m^3}$, the initial
density of the droplet is found to be $\rho_0=1,007.2\,{\rm kg/m^3}$. The corresponding initial mass of the droplet is then $m_0=\rho_0V_0$. It is noted
that the initial 98 wt\% of water content is equivalent to initial 99 vol\% of water, which is obtained by multiplying 98 with $\rho_0/\rho_{\rm w}=1007.2/997=1.01$. Consequently, the initial 2 wt\% of salt and protein together is equivalent to only 1 vol\% of their initial volume content.

Expressions (\ref{3.13}) and (\ref{3.14}) specify the mass $m_{\rm sol}^{\rm aq}=m_{\rm sol}^{\rm aq}(V_0,V)$ in terms of the known $V_0$ and $V$, and the known all other parameters that appear in these expressions. The density of the aqueous salt solution follows
from (\ref{3.6}) and is
\begin{equation}\label{3.16}
\rho_{\rm sol}^{\rm aq}=\rho_{\rm w}+\kappa\,f_{\rm s}\frac{\rho_0V_0}{m_{\rm sol}^{\rm aq}}\,,
\end{equation}
where $m_{\rm sol}^{\rm aq}$ is given by (\ref{3.13}) and (\ref{3.14}). In particular, it follows that $(\rho_{\rm sol}^{\rm aq})^0=1,004.6\,{\rm kg/m^3}$.

Finally, the total mass of the droplet is obtained by substituting (\ref{3.13}) into (\ref{3.9}). This gives
\begin{equation}\label{3.17}
m=m_{\rm sol}^{\rm aq}+m_{\rm p}=m_{\rm sol}^{\rm aq}+f_{\rm p}\rho_0V_0\,,
\end{equation}
with the corresponding density
\begin{equation}\label{3.18}
\rho=\frac{m}{V}=\frac{m_{\rm sol}^{\rm aq}}{V}+f_{\rm p}\rho_0\,\frac{V_0}{V}\,,
\end{equation}
where, again, $m_{\rm sol}^{\rm aq}$ is as specified in (\ref{3.13}) and (\ref{3.14}).

The saturation point of the aqueous salt solution is reached when $m_{\rm sol}^{\rm aq}$ decreases to the value $\hat{m}_{\rm sol}^{\rm aq}$, such that
\begin{equation}\label{3.19}
\frac{m_{\rm s}}{\hat{m}_{\rm sol}^{\rm aq}}=0.357\quad \Rightarrow \quad \hat{m}_{\rm sol}^{\rm aq}=\frac{m_{\rm s}}{0.357}=\frac{f_{\rm s}m_0}{0.357}=0.028m_0\,.
\end{equation}
The corresponding density of the saturated aqueous solution is, from (\ref{3.6}),
\begin{equation}\label{3.20}
\hat{\rho}_{\rm sol}^{\rm aq}=\rho_{\rm w}+\kappa\,\frac{m_{\rm s}}{\hat{m}_{\rm sol}^{\rm aq}}=\rho_{\rm w}+0.357\kappa=1,273.7\,{\rm kg/m^3}\,,
\end{equation}
while its volume is
\begin{equation}\label{3.21}
\hat{V}_{\rm sol}^{\rm aq}=\frac{\hat{m}_{\rm sol}^{\rm aq}}{\hat{\rho}_{\rm sol}^{\rm aq}}=0.0222V_0\,.
\end{equation}

The mass, volume, and density of the entire droplet at this instant are
\begin{equation}\label{3.22}
\begin{split}
\hat{m}&=\hat{m}_{\rm sol}^{\rm aq}+m_{\rm p}=\frac{f_{\rm s}m_0}{0.357}+f_{\rm p}m_0=0.038m_0\,,\\
\hat{V}&=\hat{V}_{\rm sol}^{\rm aq}+V_{\rm p}=\hat{V}_{\rm sol}^{\rm aq}+\frac{f_{\rm p}\rho_0V_0}{\rho_{\rm p}}=0.03V_0\,,\\
\hat{\rho}&=\frac{\hat{m}}{\hat{V}}=1.2795\rho_0=1,288.7\,{\rm kg/m^3}\,.
\end{split}
\end{equation}

The corresponding radius of the droplet and its bounding surface are
\begin{equation}\label{3.23}
\hat{R}=\left(\frac{3\hat{V}}{4\pi}\right)^{1/3}=0.31R_0\,,\quad \hat{S}=4\pi \hat{R}^2=0.096S_0\,.
\end{equation}
Consequently, from (\ref{3.5}), the time at the instant of the saturation is
\begin{equation}\label{3.24}
\hat{t}=\frac{S_0-\hat{S}}{s}=\frac{0.904S_0}{s}=0.904\tau\,,\quad \tau=S_0/s\,.
\end{equation}
The parameter $s$ depends on relative humidity as defined in (\ref{3.4}).

\subsection{Density expressions after saturation of aqueous salt solution}

As the water continues to evaporate beyond the saturation time $\hat{t}$ given by (\ref{3.22}), a portion of the total salt amount $m_{\rm s}$
(denoted by $m_{\rm s}''$) settles at the bottom of the saturated aqueous solution. The remaining portion of salt
$m_{\rm s}'=0.357m_{\rm sol}^{\rm sat}$ is dissolved in the saturated aqueous solution, and $m_{\rm s}=m_{\rm s}'+m_{\rm s}''$.
The total mass of the droplet at this stage of evaporation is, thus,
\begin{equation}\label{3.25}
m=m_{\rm sol}^{\rm sat}+m_{\rm s}''+m_{\rm p}=m_{\rm sol}^{\rm sat}+(m_{\rm s}-m_{\rm s}')+m_{\rm p}=0.643m_{\rm sol}^{\rm sat}+m_{\rm s}+m_{\rm p}\,.
\end{equation}
Furthermore, the volume of the droplet is
\begin{equation}\label{3.26}
V=V_{\rm sol}^{\rm sat}+V_{\rm s}''+V_{\rm p}=\frac{m_{\rm sol}^{\rm sat}}{\hat{\rho}_{\rm sol}^{\rm aq}}+\frac{m_{\rm s}''}{\rho_{\rm s}}+\frac{m_{\rm p}}{\rho_{\rm p}}=
\frac{m_{\rm sol}^{\rm sat}}{\hat{\rho}_{\rm sol}^{\rm aq}}+\frac{m_{\rm s}-0.357m_{\rm sol}^{\rm sat}}{\rho_{\rm s}}+\frac{m_{\rm p}}{\rho_{\rm p}}\,,
\end{equation}
where $\hat{\rho}_{\rm sol}^{\rm aq}=1,273.7\,{\rm kg/m^3}$, as determined in (\ref{3.20}), and $\rho_{\rm s}=2,160\,{\rm kg/m^3}$ is the density of solid (crystalline) NaCl.
Equation (\ref{3.26}) can be solved for $m_{\rm sol}^{\rm sat}$ to obtain
\begin{equation}\label{3.27}
m_{\rm sol}^{\rm sat}=\hat{\rho}_{\rm sol}^{\rm aq}\,\cfrac{V-m_{\rm s}/\rho_{\rm s}-m_{\rm p}/\rho_{\rm p}}{1-0.357\,\hat{\rho}_{\rm sol}^{\rm aq}/\rho_{\rm s}}\,,
\end{equation}
where $m_{\rm s}=f_{\rm s}m_0$, $m_{\rm p}=f_{\rm p}m_0$, and $m_0=\rho_0V_0$. With the numerical values substituted, (\ref{3.27}) becomes
\begin{equation}\label{3.28}
m_{\rm sol}^{\rm sat}=\alpha_1V-\beta_1V_0\,,\quad \alpha_1=1,613\,{\rm kg/m^3}\,,\quad \beta_1=19.559\,{\rm kg/m^3}\,.
\end{equation}
By substituting (\ref{3.27}) into (\ref{3.25}), we obtain
\begin{equation}\label{3.29}
m=\alpha V+\beta V_0\,,
\end{equation}
where
\begin{equation}\label{3.30}
\alpha=\cfrac{0.643\hat{\rho}_{\rm sol}^{\rm aq}}{1-0.357\,\hat{\rho}_{\rm sol}^{\rm aq}/\rho_{\rm s}}\,,\quad
\beta=\rho_0\left(f_{\rm s}+f_{\rm p}-0.643\,\hat{\rho}_{\rm sol}^{\rm aq}\,\cfrac{f_{\rm s}/\rho_{\rm s}+f_{\rm p}/\rho_{\rm p}}{1-0.357\,\hat{\rho}_{\rm sol}^{\rm aq}/\rho_{\rm s}}\right).
\end{equation}
The numerical values for $\alpha$ and $\beta$ are $\alpha=1,030\,{\rm kg/m^3}$ and $\beta=7.657\,{\rm kg/m^3}$.
Finally, the density of the droplet during this stage of its evaporation is, from (\ref{3.29}),
\begin{equation}\label{3.31}
\rho=\frac{m}{V}=\alpha +\beta\,\frac{V_0}{V}\,.
\end{equation}

\subsection{Size of the droplet's residue}

If all water evaporates and only salt and dry protein remain in their original amount, the mass of the droplet's residue is $m_*=m_{\rm s}+m_{\rm p}$, as given in (\ref{3.7}).
(Proteins in aqueous solutions bind some of the water molecules very firmly, but in our model we consider that all the droplet's water can eventually evaporate during the droplet's motion in the air.) According to (\ref{3.29}), this must be equal to $\alpha V_*+\beta V_0$, where $V_*$ is the volume of the droplet's residue, i.e.,
\begin{equation}\label{3.32}
\alpha V_*+\beta V_0=(f_{\rm s}+f_{\rm p})\rho_0V_0\,.
\end{equation}
Thus,
\begin{equation}\label{3.33}
V_*=\frac{1}{\alpha}\,[(f_{\rm s}+f_{\rm p})\rho_0-\beta]V_0=0.0121V_0\,.
\end{equation}
The average density of the droplet's residue is
\begin{equation}\label{3.34}
\rho_*=\frac{m_*}{V_*}=\frac{(f_{\rm s}+f_{\rm p})\rho_0\alpha}{(f_{\rm s}+f_{\rm p})\rho_0-\beta}=1.6496\rho_0=1,661.5\,{\rm kg/m^3}\,,
\end{equation}
while the corresponding radius is
\begin{equation}\label{3.35}
R_*=\left(\frac{2V_*}{4\pi}\right)^{1/3}=\sqrt[3]{0.0121}\,R_0=0.23R_0\,,\quad S_*=4\pi R_*^2=0.6625S_0\,.
\end{equation}
The time at the instant when the droplet reaches its residue (nucleus) size is
\begin{equation}\label{3.36}
t_*=\frac{S_0-S_*}{s}=0.3375\tau\,,\quad \tau=S_0/s\,.
\end{equation}

We note that the residue's density can also be derived directly from the additive mass and volume expressions
\begin{equation}\label{3.37}
\begin{split}
m_*&=\rho_*V_*=m_{\rm s}+m_{\rm p}=(f_{\rm s}+f_{\rm p})m_0\,,\\
V_*&=V_{\rm s}+V_{\rm p}=\frac{m_{\rm s}}{\rho_{\rm s}}+\frac{m_{\rm p}}{\rho_{\rm p}}=\left(\frac{f_{\rm s}}{\rho_{\rm s}}+\frac{f_{\rm p}}{\rho_{\rm p}}\right)\!\rho_0V_0,
\end{split}
\end{equation}
from which it follows that the density of the residue particle is
\begin{figure}
\centering{
\includegraphics[scale=0.45]{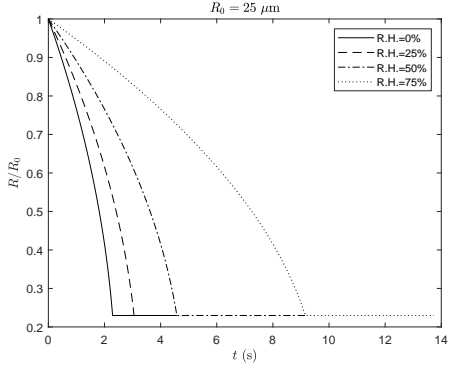}\hskip4mm
\includegraphics[scale=0.45]{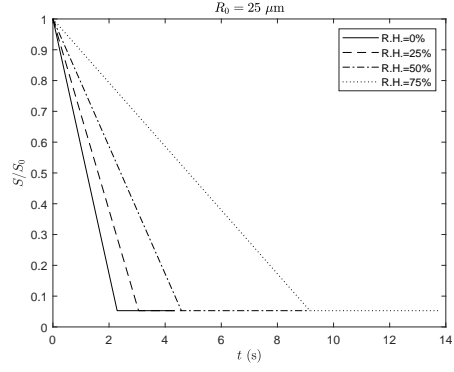}
}
\vskip3mm
\centerline{(a)\hskip80mm (b)}
\vskip3mm
\centering{
\includegraphics[scale=0.45]{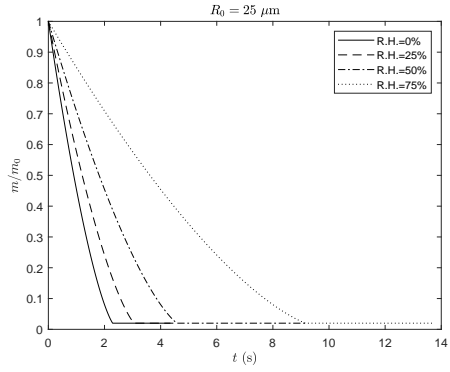}\hskip4mm
\includegraphics[scale=0.45]{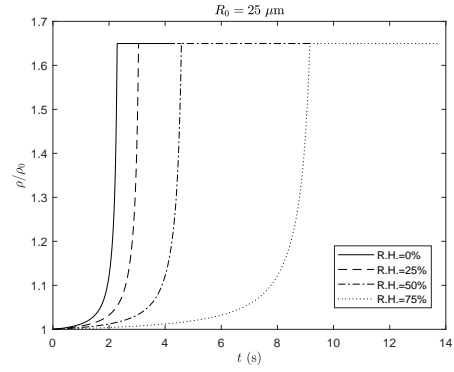}
}
\vskip3mm
\centerline{(c)\hskip80mm (d)}
\caption{\label{Fig2} The variation of the (a) radius, (b) surface, (c) mass, and (d) density of the evaporating droplet with initial radius $R_0=25\mu$m under four different relative humidities.}
\end{figure}
\begin{equation}\label{3.38}
\rho_*=\frac{m_*}{V_*}=\frac{m_{\rm s}+m_{\rm p}}{m_{\rm s}/\rho_{\rm s}+m_{\rm p}/\rho_{\rm p}}=\frac{f_{\rm s}+f_{\rm p}}{f_{\rm s}/\rho_{\rm s}+f_{\rm p}/\rho_{\rm p}}\,.
\end{equation}

\section{Numerical results}

Figure \ref{Fig2} shows the variations of the radius, volume,  mass, and density of the evaporating droplet of initial radius $R_0=25\, \mu$m under four indicated values of relative humidity.
The initial mass of the droplet is $m_0=6.592\times 10^{-11}$ kg and its average density is $\rho_0=1,007.2\,{\rm kg/m^3}$. The initial mass of aqueous salt solution of the droplet is $(m_{\rm sol}^{\rm aq})^0=6.526\times 10^{-11}$ kg and its density $(\rho_{\rm sol}^{\rm aq})^0=1,004.6\,{\rm kg/m^3}$. When aqueous salt solution becomes saturated, its mass, volume, and density are $\hat{m}_{\rm sol}^{\rm aq}=1.8465\times 10^{-12}$ kg, $\hat{V}_{\rm sol}^{\rm aq}=1.458\times 10^{-15}\,{\rm m^3}$,
and $\hat{\rho}_{\rm sol}^{\rm aq}=1,266.5\,{\rm kg/m^3}$, respectively. The corresponding radius, mass, volume, and density of the entire droplet at that instant
are $\hat{R}=0.31R_0=7.745\,\mu$m, $\hat{m}=0.038m_0=2.506\times 10^{-12}$ kg, $\hat{V}=0.03V_0=1.946\times 10^{-15}\,{\rm m^3}$,
and $\hat{\rho}=1.278\rho_0=1,287.5\,{\rm kg/m^3}$. At the instant when all water evaporates and the droplet reduces to its residue size, its radius is $R_*=0.23R_0=5.743\,\mu$m. The corresponding mass, volume, and density are
$m_*=0.02m_0=1.318\times 10^{-12}$ kg, $V_*=0.0121V_0=7.935\times 10^{-16}\,{\rm m^3}$,  and $\rho_*=1.6496\rho_0=1,661.5\,{\rm kg/m^3}$.

The times to reach the aqueous salt solution saturation state and the droplet's residue state depend on relative humidity. For relative humidities of 0\%, 25\%, 50\%, 75\%, these times are $\hat{t}=$2.18, 2.91, 4.37, 8.74 seconds and  $t_*=$2.29, 3.05, 4.58, 9.16 seconds, respectively. These values are for $R_0=25\, \mu$m; the results for other values of the initial radius follow similarly. The times $\hat{t}$ and $t_*$ increase with the increase of the initial radius. For example, for a droplet of initial radius $R_0=50\,\mu$m, the times are
$\hat{t}=$8.74, 11.65, 17.48, 34.96 seconds and  $t_*=$9.16, 12.21, 18.31, 36.62 seconds, respectively. Depending on the height from which the droplet begins its fall, these times may be much longer than the times it takes a droplet to reach the ground. This is discussed below.

\begin{figure}
\centering{
\includegraphics[scale=0.45]{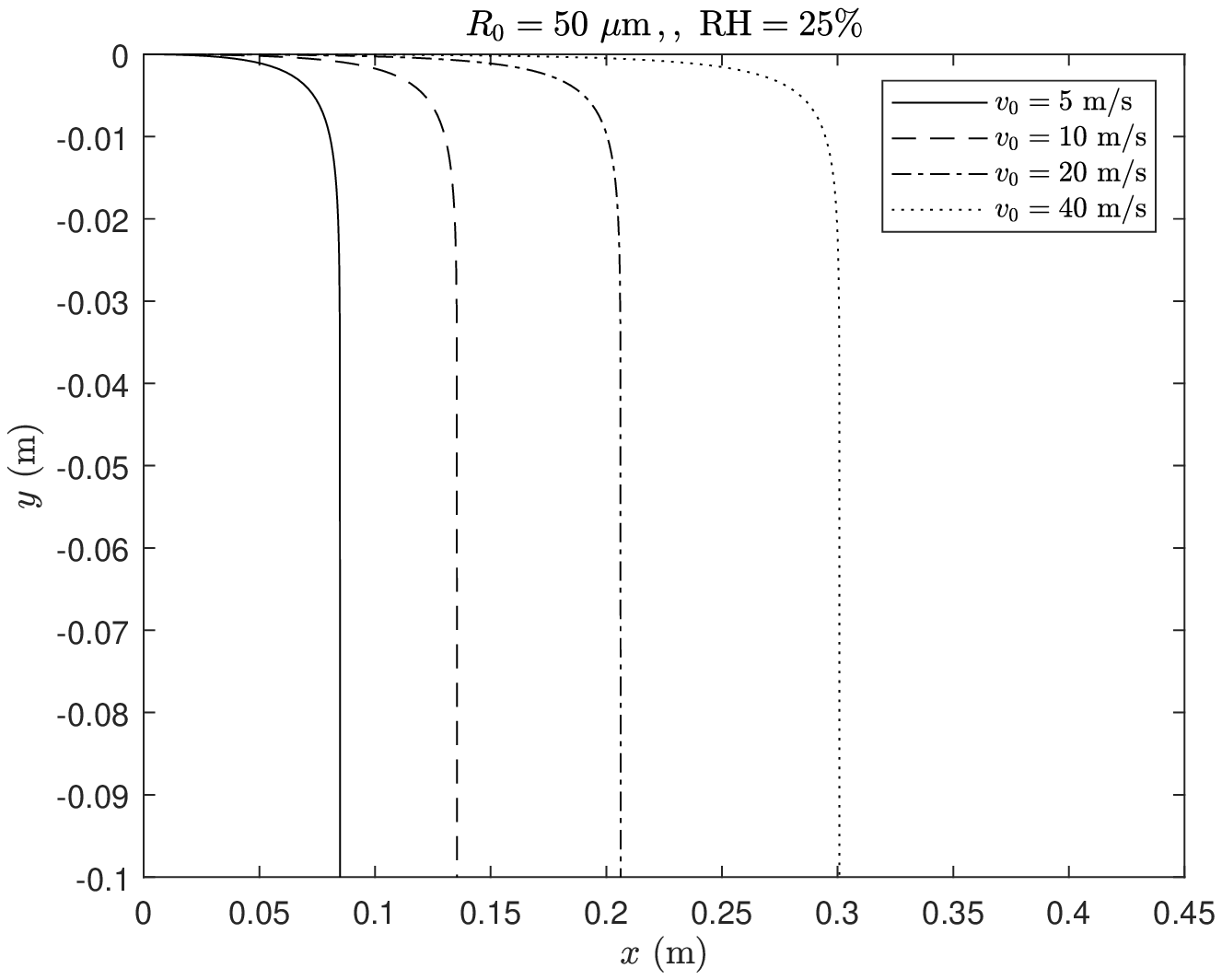}\hskip4mm
\includegraphics[scale=0.45]{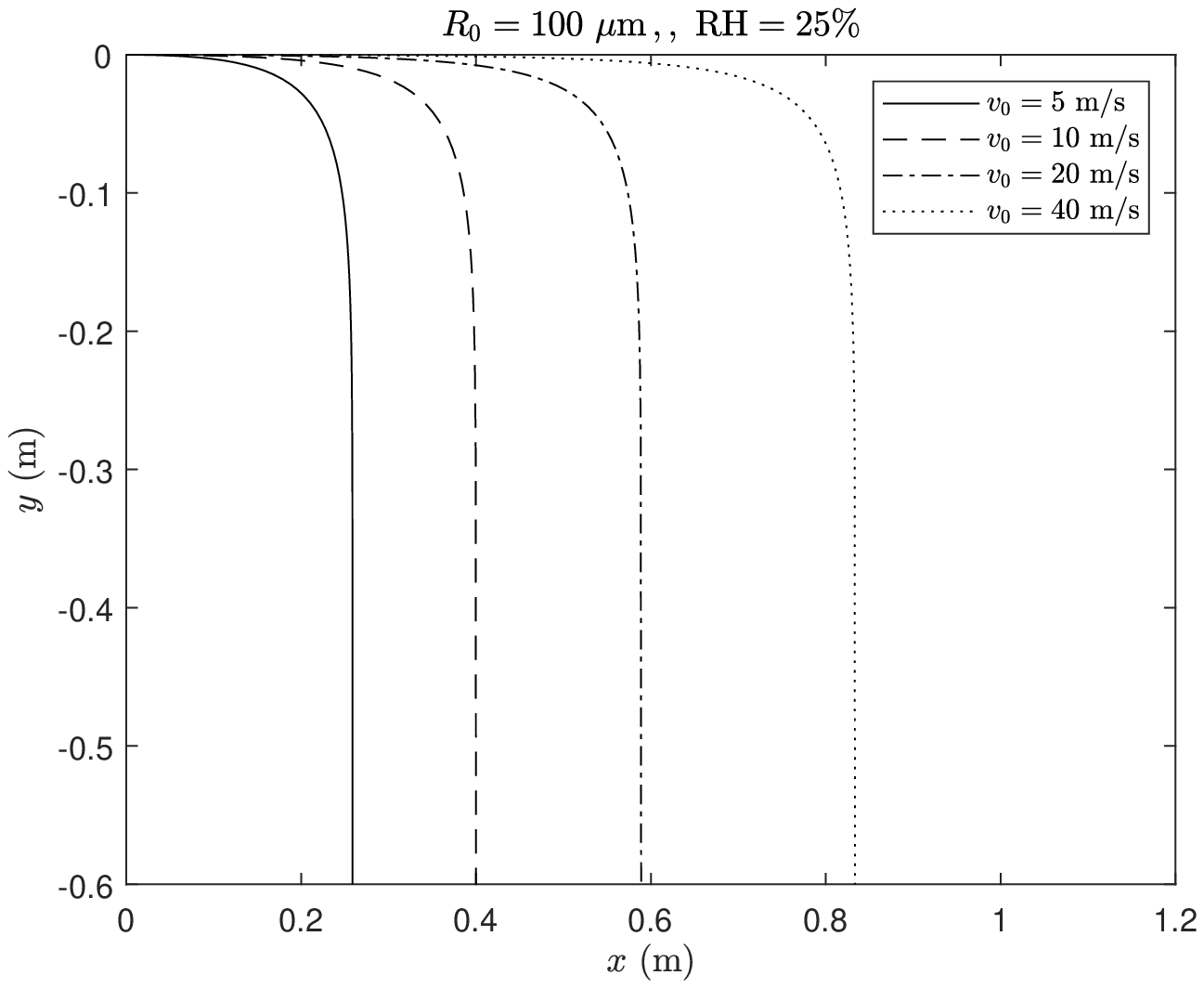}
}
\vskip3mm
\centerline{(a)\hskip80mm (b)}
\caption{\label{Fig3} The trajectories of the droplets of initial radii $R_0=50$ and 100 $\mu$m, ejected horizontally with the velocities $v_0=5$, 10, 20 and 40 m/s under relative humidity of 25\%. Upon reaching its maximum horizontal distance, the droplet falls vertically downwards.}
\end{figure}
\begin{figure}
\centering{
\includegraphics[scale=0.45]{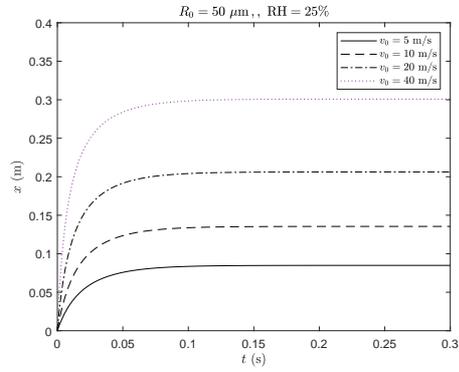}\hskip4mm
\includegraphics[scale=0.45]{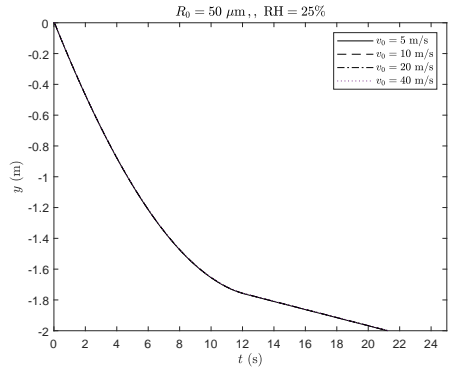}
}
\vskip3mm
\centerline{(a)\hskip80mm (b)}
\vskip3mm
\centering{
\includegraphics[scale=0.45]{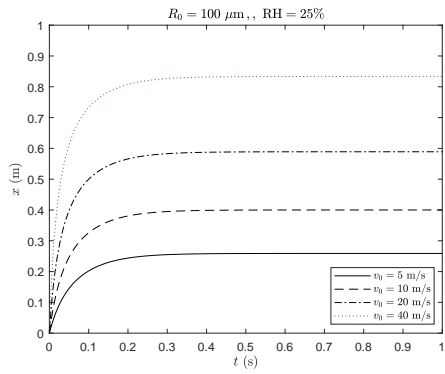}\hskip4mm
\includegraphics[scale=0.45]{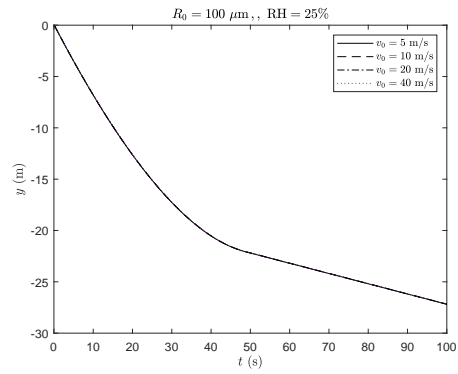}
}
\vskip3mm
\centerline{(c)\hskip80mm (d)}
\caption{\label{Fig4}  The variation of the $x$ and $y$ coordinates of the droplet's trajectory with time $t$ for droplets of initial radii $R_0=50$ (parts a and b), and  100 $\mu$m (parts c and d), ejected horizontally with the velocities $v_0=5$, 10, 20 and 40 m/s under relative humidity of 25\%.}
\end{figure}
Figure \ref{Fig3} shows the trajectories of the droplets of initial radii $R_0=50$ and 100 $\mu$m, emitted from a mouth horizontally with initial velocities $v_0=5$, 10, 20, and 40 m/s, in a stagnant air under relative humidity 25\%. Figure \ref{Fig4} shows the plots $x=x(t)$ and $y=y(t)$ for droplets with initial radii $R_0=50$ and 100 $\mu$m under the same relative humidity of 25\%. The maximum horizontal distance is rapidly reached, which is followed by essentially vertical descent of the droplet. The maximum horizontal distance is strongly
dependent on the magnitude of the initial horizontal velocity $v_0$, while the time variation $y=y(t)$ is only mildly affected by $v_0$.

For the 50 $\mu$m initial radius, the droplet evaporates to its 11.5 $\mu$m radius residue in  about 12.2 seconds. This is followed by essentially vertical fall of a non-evaporating  residue toward the ground (Fig. \ref{Fig4}a), its velocity approaching the terminal velocity $(2/9)R_*^2\rho g/\eta_{\rm air}=2.62$ cm/s. If the ground is 2 m below the point of exhalation,
then the droplet reaches the ground in about 21 seconds, i.e., about 9.5 seconds after it reaches its residue (nucleus) size.
On the other hand, the 100 $\mu$m initial radius droplet is sufficiently large to reach the ground, 2 m below the point of exhalation, in
just 3 seconds, which is far shorter than 49 seconds that would be needed to evaporate to its residue size of about 23 $\mu$m radius (Fig. \ref{Fig4}d).

Figures \ref{Fig3} and \ref{Fig4} also quantify how much further the droplet moves horizontally upon coughing ($v_0=40$ m/s) comparing to soft talking ($v_0=5$ m/s).
For example, the maximum horizontal reach of the droplet with initial radius $100\,\mu$m is about 83 cm in the case $v_0=40$ m/s, while it is just about 26 cm in the case $v_0=5$ m/s (both at relative humidity of 25\%). For the initial droplet's radius of $50\,\mu$m, the maximum horizontal reach is about 30 cm in the case $v_0=40$ m/s, while it is less than 1/3 of that (8.5 cm) in the case $v_0=5$ m/s.

\begin{figure}
\centering{
\includegraphics[scale=0.45]{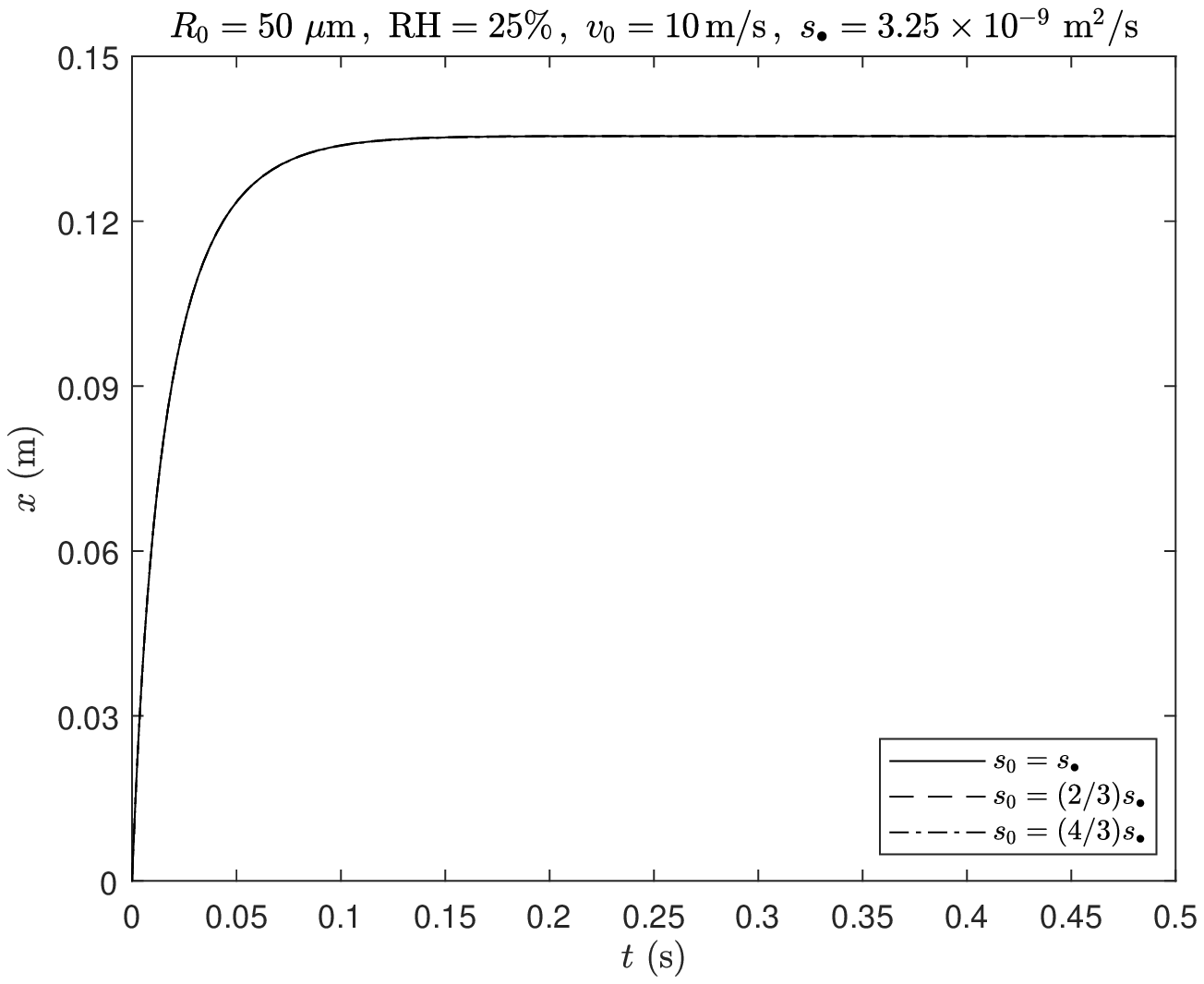}\hskip4mm
\includegraphics[scale=0.45]{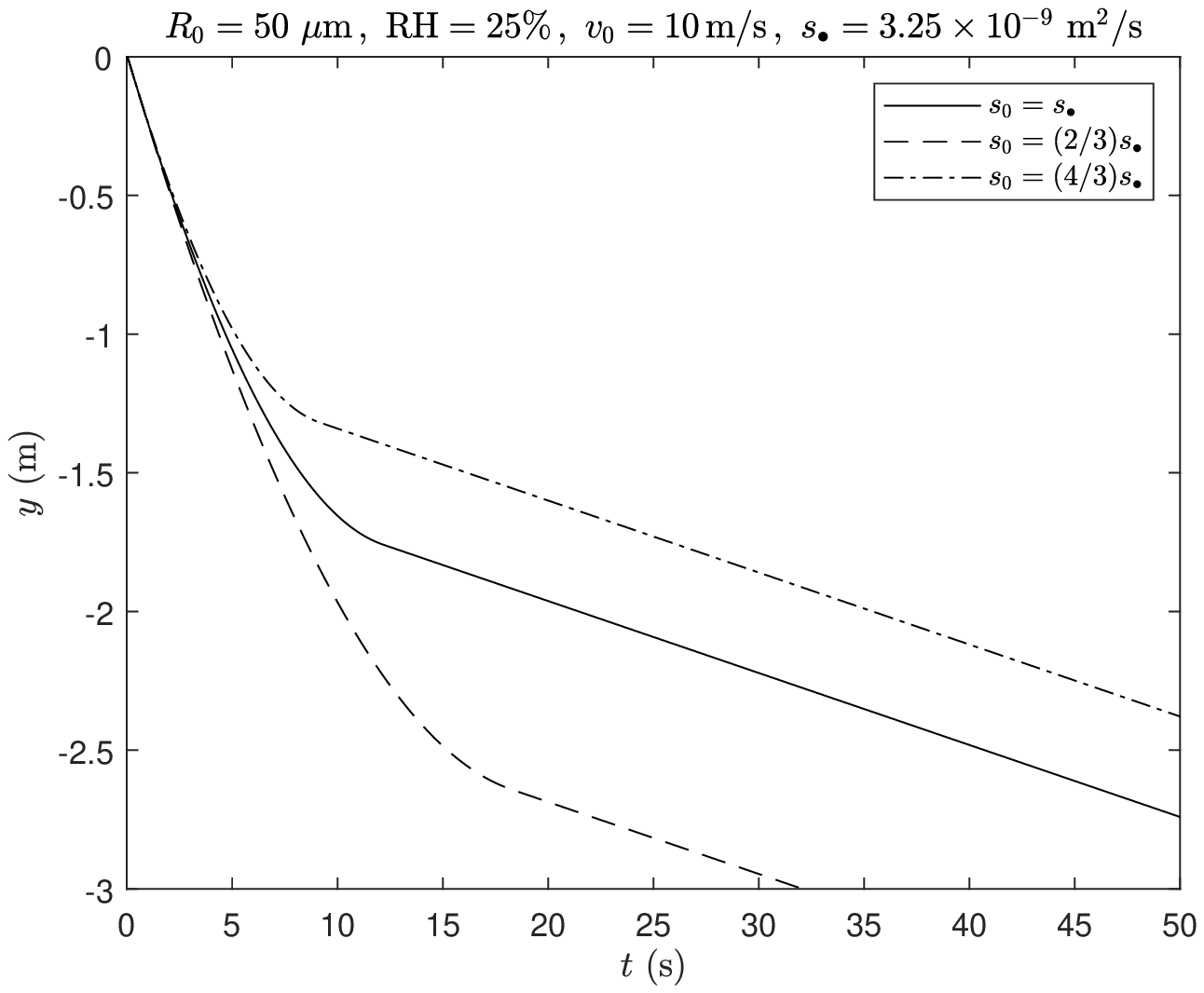}
}
\vskip3mm
\centerline{(a)\hskip80mm (b)}
\caption{\label{Fig5} The effect of the rate of evaporation $s_0$ on the time variation of the droplet's $(x,y)$ coordinates. (a) The $x$ coordinate is very little affected by the change of $s_0$. (b) The $y$ coordinate is affected substantially as the droplet descends vertically and reaches its nucleus size much sooner for higher values of $s_0$.}
\end{figure}
\begin{figure}
\centering{
\includegraphics[scale=0.45]{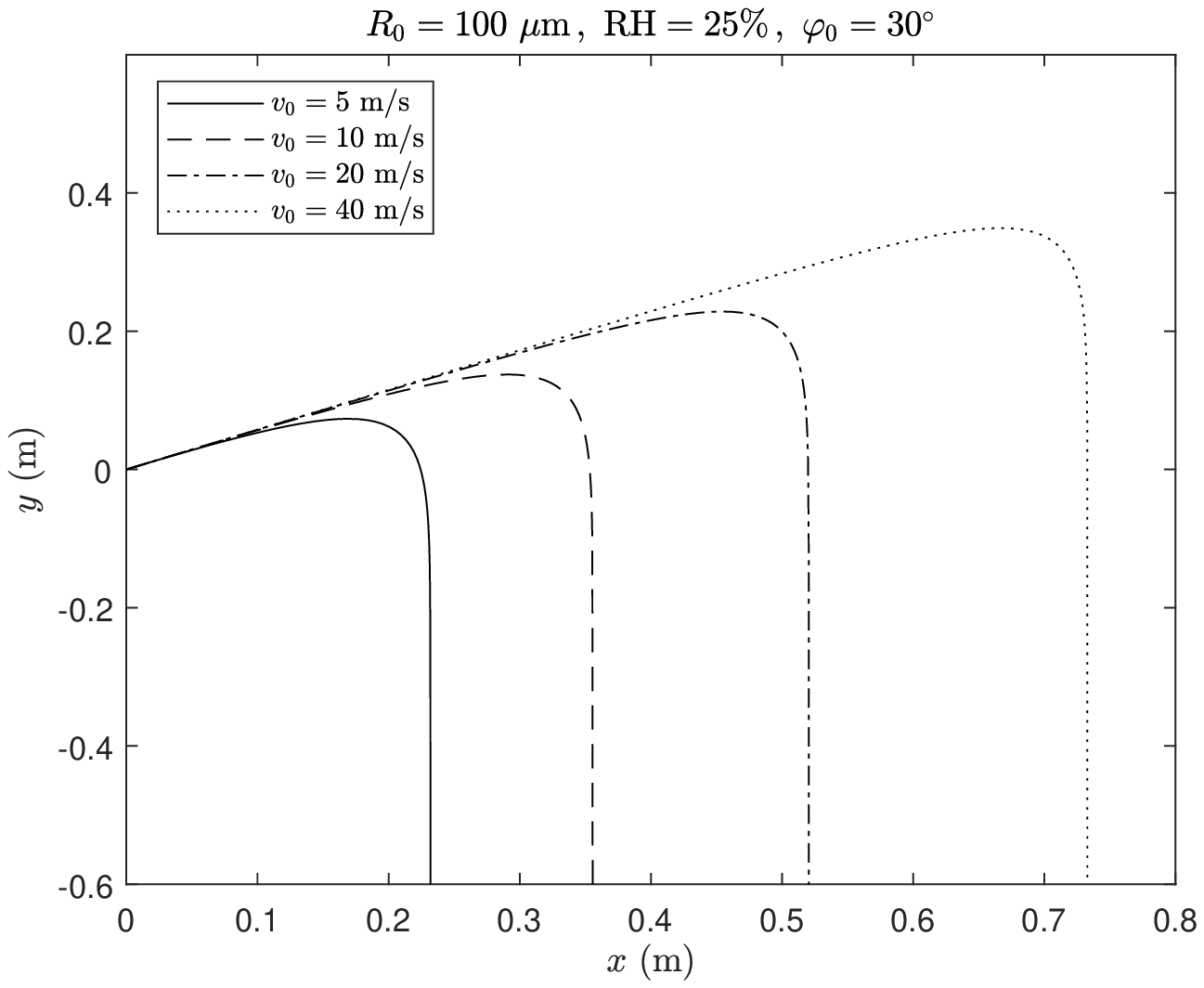}\hskip4mm
\includegraphics[scale=0.45]{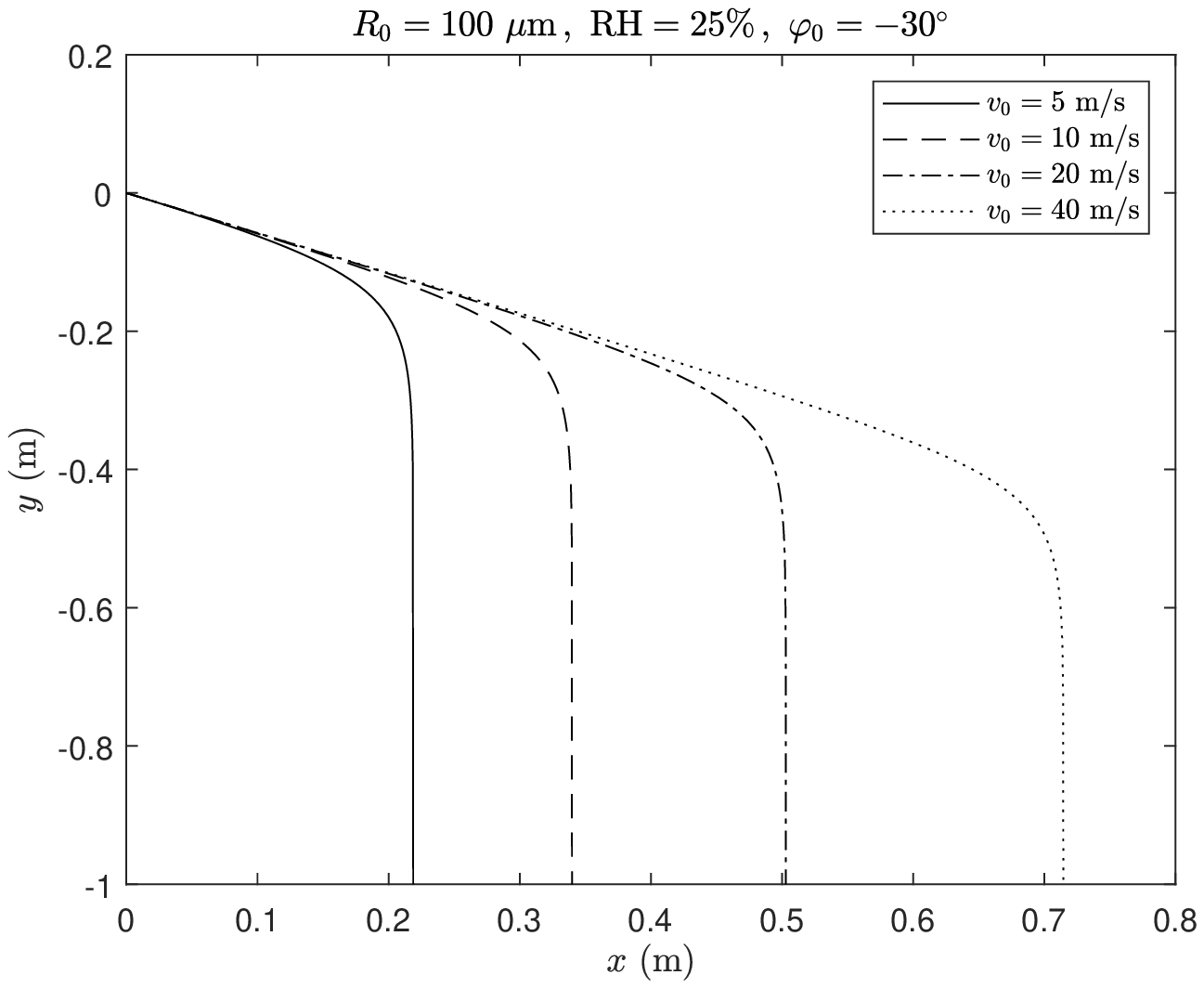}
}
\vskip3mm
\centerline{(a)\hskip80mm (b)}
\vskip3mm
\centering{
\includegraphics[scale=0.45]{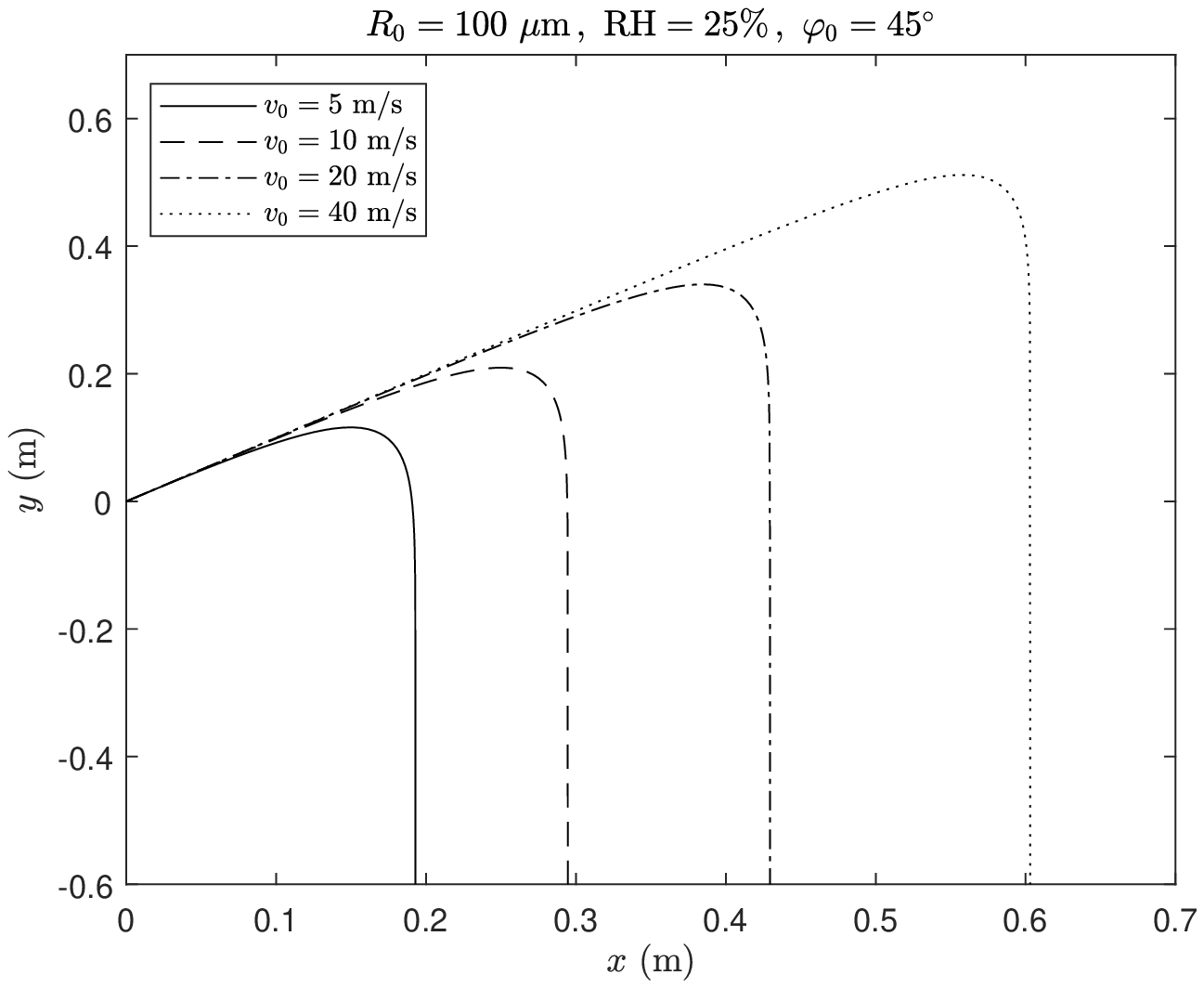}\hskip4mm
\includegraphics[scale=0.45]{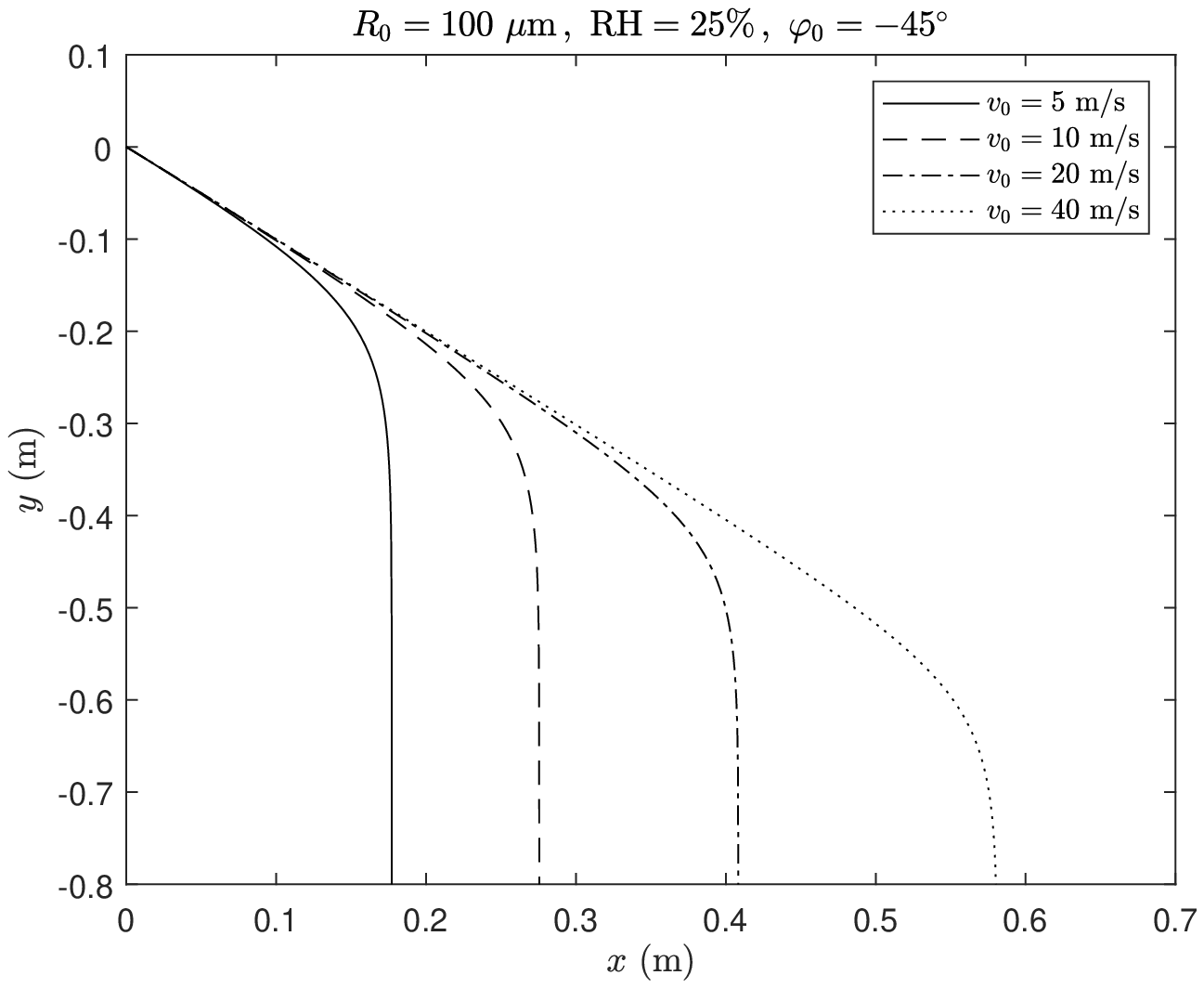}
}
\vskip3mm
\centerline{(c)\hskip80mm (d)}
\caption{\label{Fig6} The trajectories of the droplet of initial radius 100 $\mu$m, ejected from mouth with the velocities $v_0=5$, 10, 20 and 40 m/s under relative humidity of 25\% at the angles $\varphi_0=\pm 30^\circ$ and $\pm 45^\circ$.}
\end{figure}
Figure \ref{Fig5} shows the effect of the rate of evaporation $s_0$ on the time-variation of the droplet's position. Three values of $s_0$ are used in the plots, $s_0=s_\bullet=3.25\times
10^{-9}\,{\rm m^2/s}$, $s_0=(2/3)s_\bullet$, and $s_0=(4/3)s_\bullet$. The horizontal position ($x$ coordinate) is very little affected by $s_0$, as the droplet rapidly looses (within less than 0.2 seconds) its horizontal momentum due to the air drag, before any substantial evaporation takes place. The vertical position ($y$ coordinate) is, however, affected substantially
by the change of $s_0$, as the droplet descends vertically by the action of gravity, while its size and weight substantially decrease due to evaporation. The droplet reaches its nucleus size (the kink in the curves shown in Fig. \ref{Fig5}) much sooner for higher values of $s_0$. The time it takes for the droplet of initial radius $50\,\mu$m, ejected at the speed of $10\,$ m/s, to descend by 2 m is about 10 seconds in the case $s_0=(2/3)s_\bullet$,  21 seconds in the case $s_0=s_\bullet$, and 35 seconds in the case $s_0=(4/3)s_\bullet$. The relative humidity in all three cases is assumed to be the same (R.H.=25\%). Thus, at higher ambient temperature (higher value of $s_0$), the droplet remains in air at higher hight longer, which increases the risk of infection (more rapidly evaporating droplets are lighter and fall to the ground slower).

Finally, Fig. \ref{Fig6} shows the trajectories of the droplet of initial radius 100 $\mu$m, ejected with initial velocities $v_0=5$, 10, 20 and 40 m/s under relative humidity of 25\% at the angles $\varphi_0=\pm 30^\circ$ and $\pm 45^\circ$. The results illustrate the effect of the initial angle on the horizontal reach of the droplet. This is of importance because a person can, for example, speak while sitting in the chair next to the person standing by the chair, or vice versa. Clearly, the maximum reach increases with the decrease of the magnitude of the angle $\varphi_0$, being the greatest for $\varphi_0=0^\circ$. These numerical results are of importance for the evaluation of the risks of infection by transmitted pathogenic droplets
and the estimates of safe distancing while communicating during standing and/or sitting, or from different elevations and heights. For example, the maximum vertical reach of a 100 $\mu$m-radius droplet ejected by soft talking with $v_0=5$ m/s at $\varphi_0=45^\circ$ is $y_{\rm max}=11.6$ cm, while its maximum horizontal reach is $x_{\rm max}=19.3$ cm. On the other hand, the maximum vertical reach of the same droplet ejected by coughing with $v_0=40$ m/s is $y_{\rm max}=51.2$ m, while its maximum horizontal reach is $x_{\rm max}=60.3$ cm. If the droplet was ejected horizontally $(\varphi_0=0^\circ$), the maximum horizontal reach would have been $25.9$ cm in soft talking and $83.4$ cm in coughing (Fig. \ref{Fig3}d).
\begin{figure}
\centering{
\includegraphics[scale=0.45]{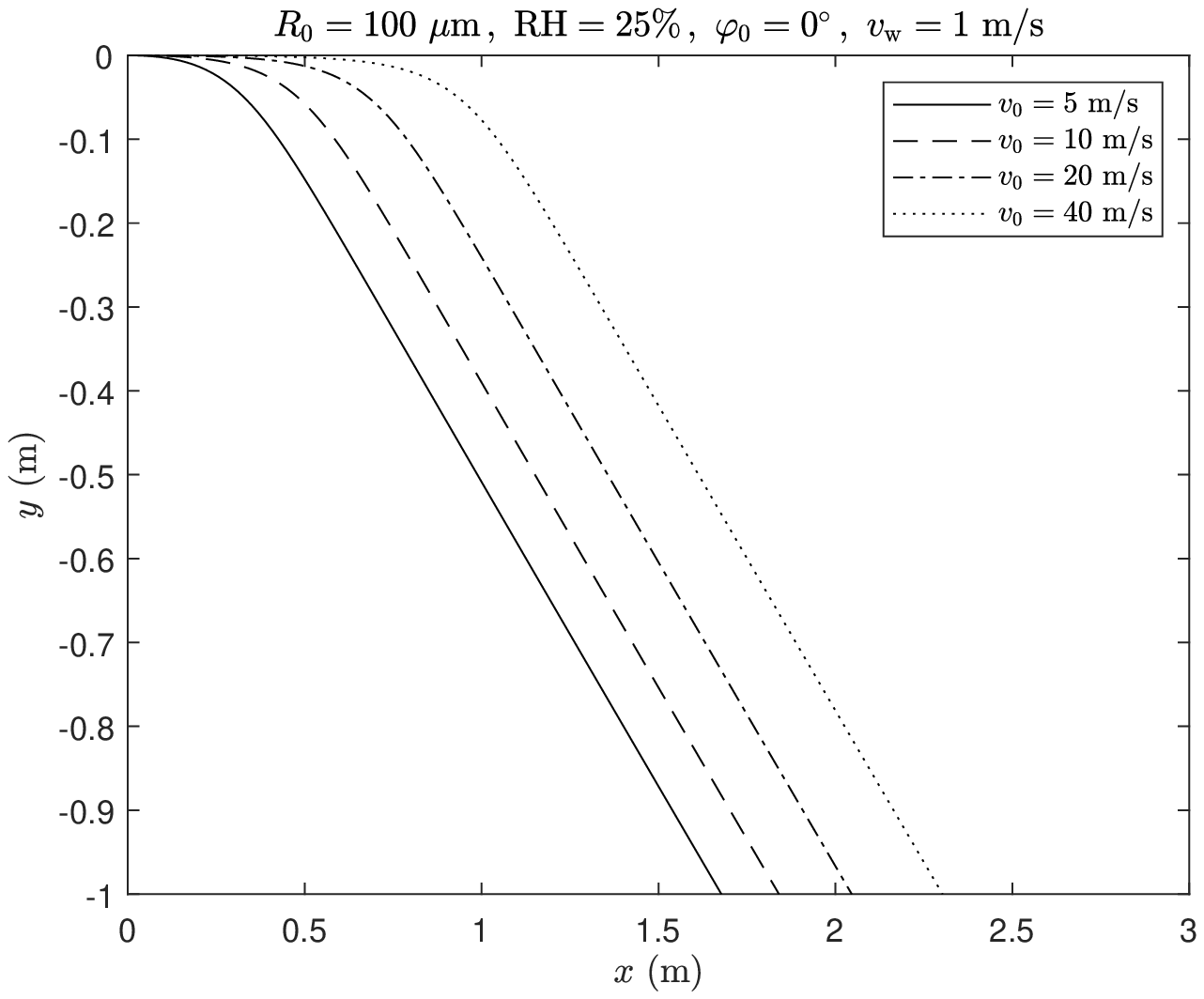}\hskip4mm
\includegraphics[scale=0.45]{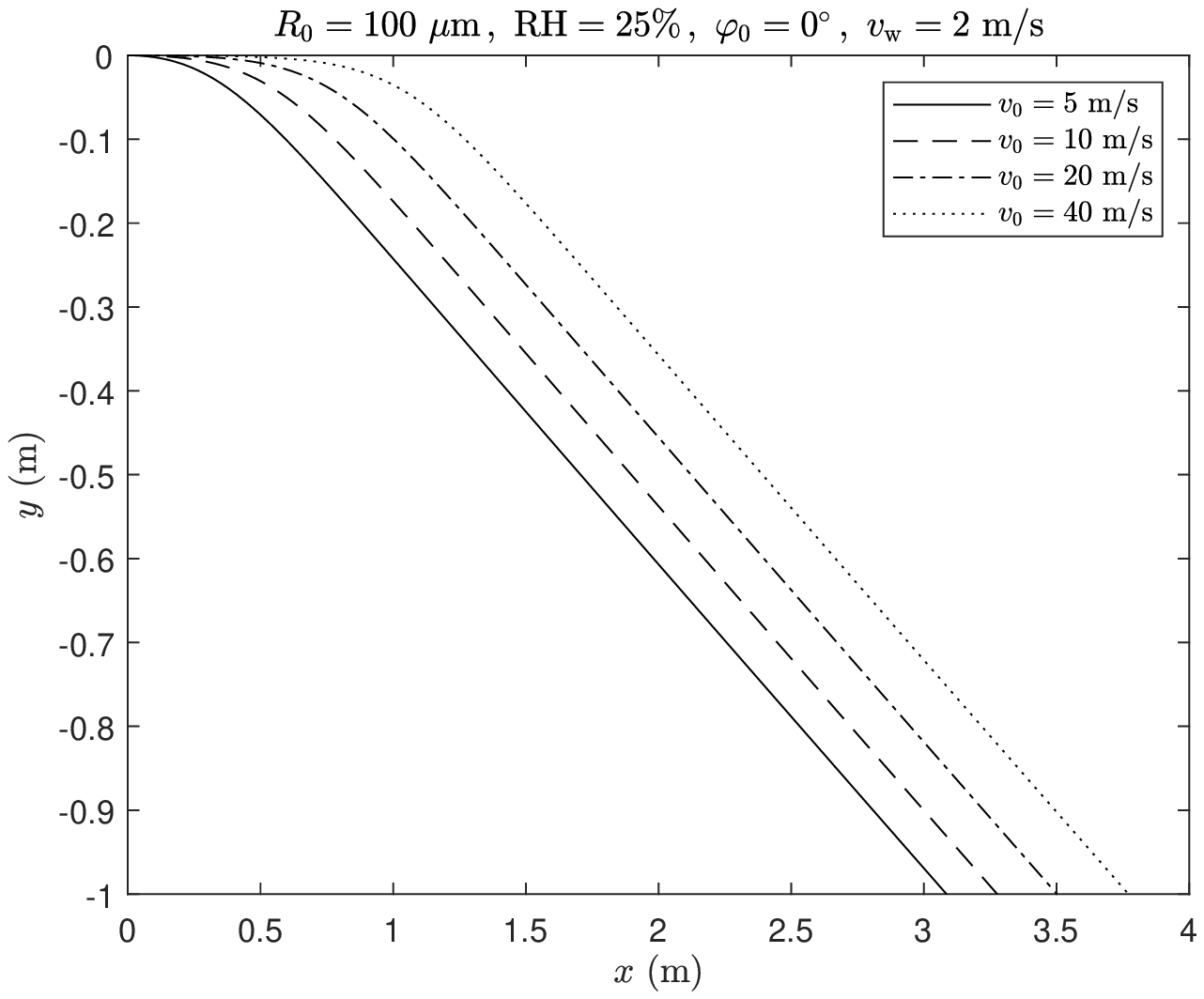}
}
\vskip3mm
\centerline{(a)\hskip80mm (b)}
\vskip3mm
\centering{
\includegraphics[scale=0.45]{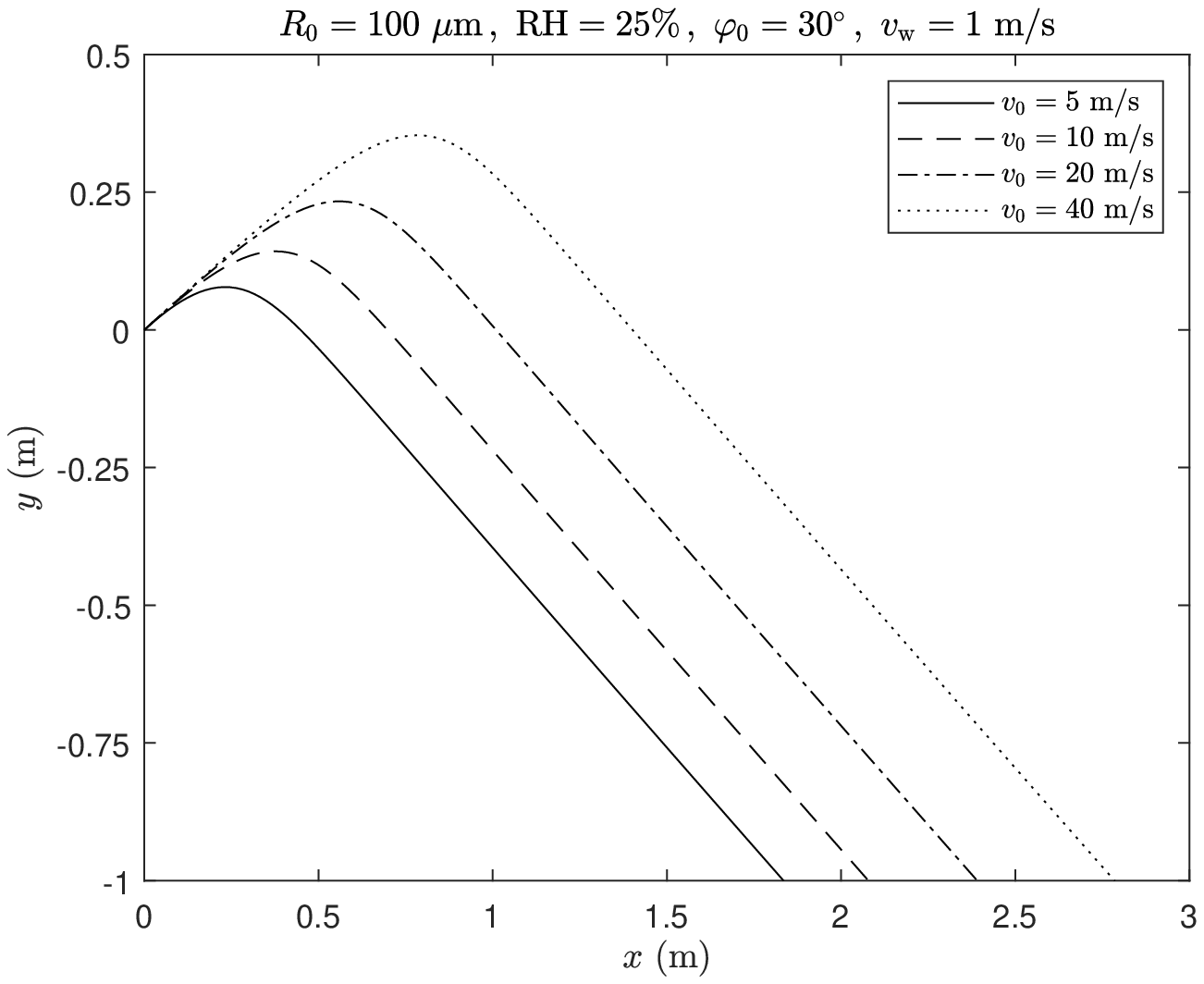}\hskip4mm
\includegraphics[scale=0.45]{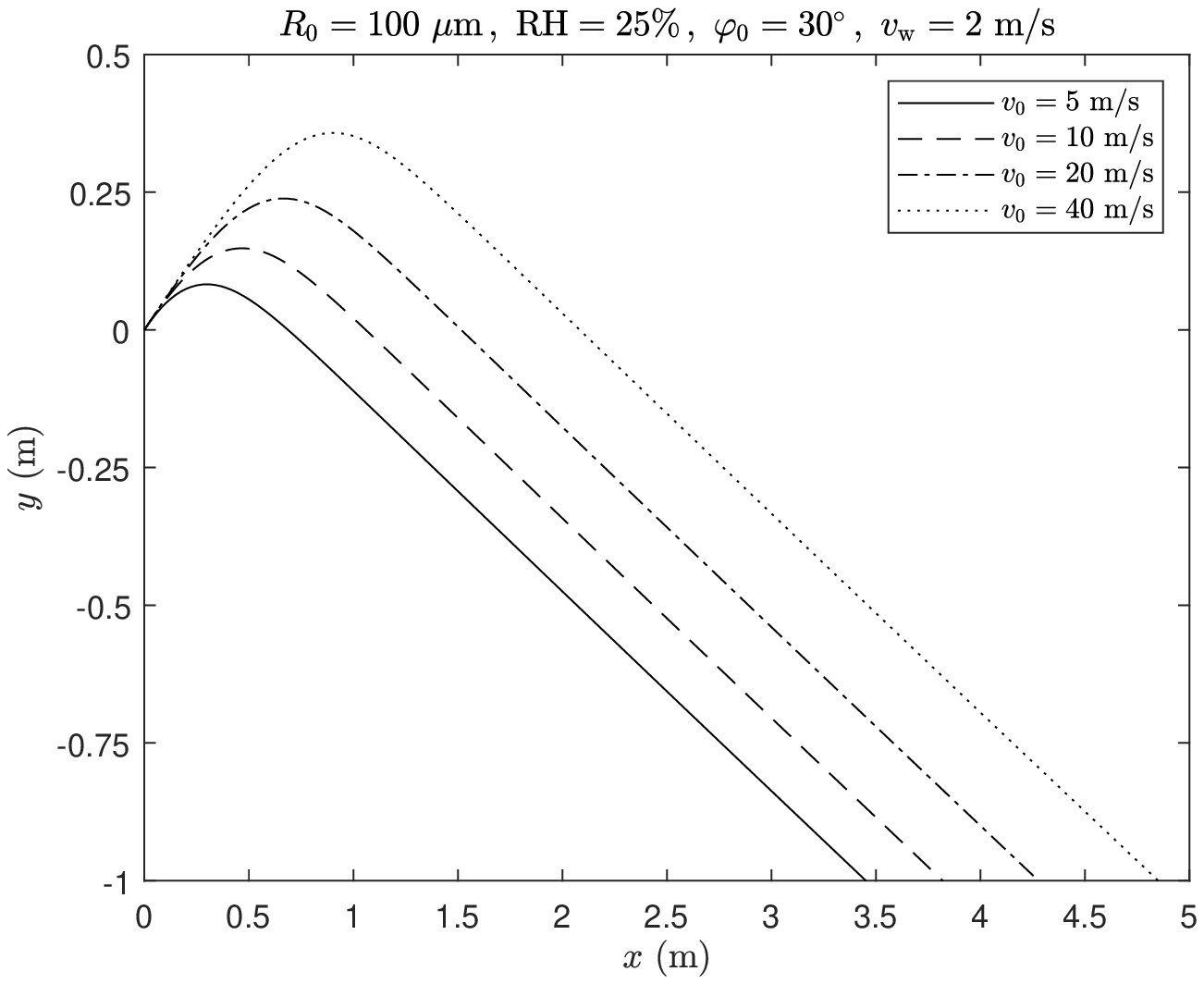}
}
\vskip3mm
\centerline{(c)\hskip80mm (d)}
\vskip2mm
\caption{\label{Fig7} The trajectories of the droplet of initial radius 100 $\mu$m, ejected with the velocities $v_0=5$, 10, 20 and 40 m/s at the angle $\varphi_0=0$ in parts (a) and (b), and $\varphi_0=30^\circ$ in parts (c) and (d). The wind velocity in parts (a) and (c) is 1 m/s, and 2 m/s in parts (b) and (d).}
\end{figure}
\begin{figure}
\centering{
\includegraphics[scale=0.45]{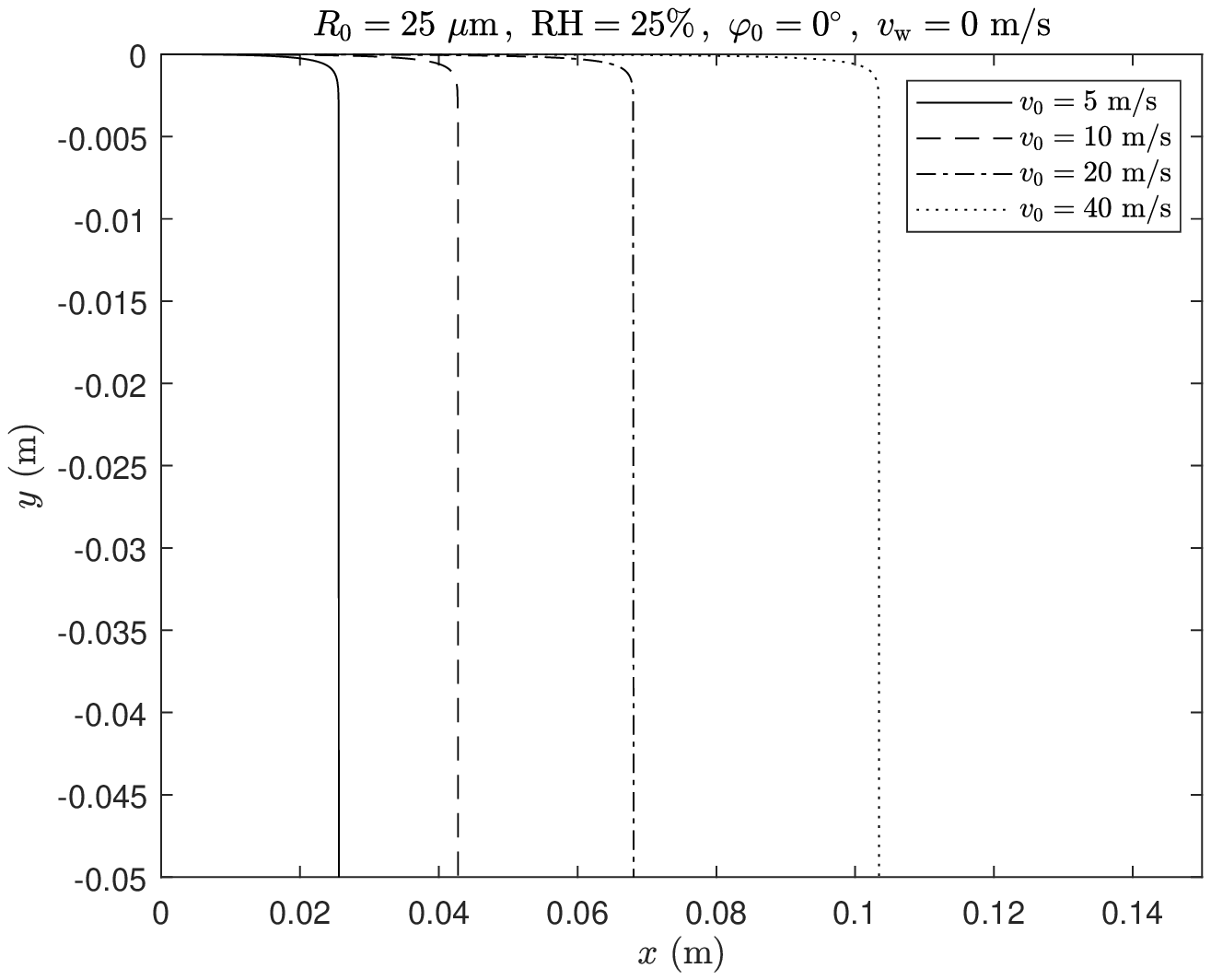}\hskip4mm
\includegraphics[scale=0.45]{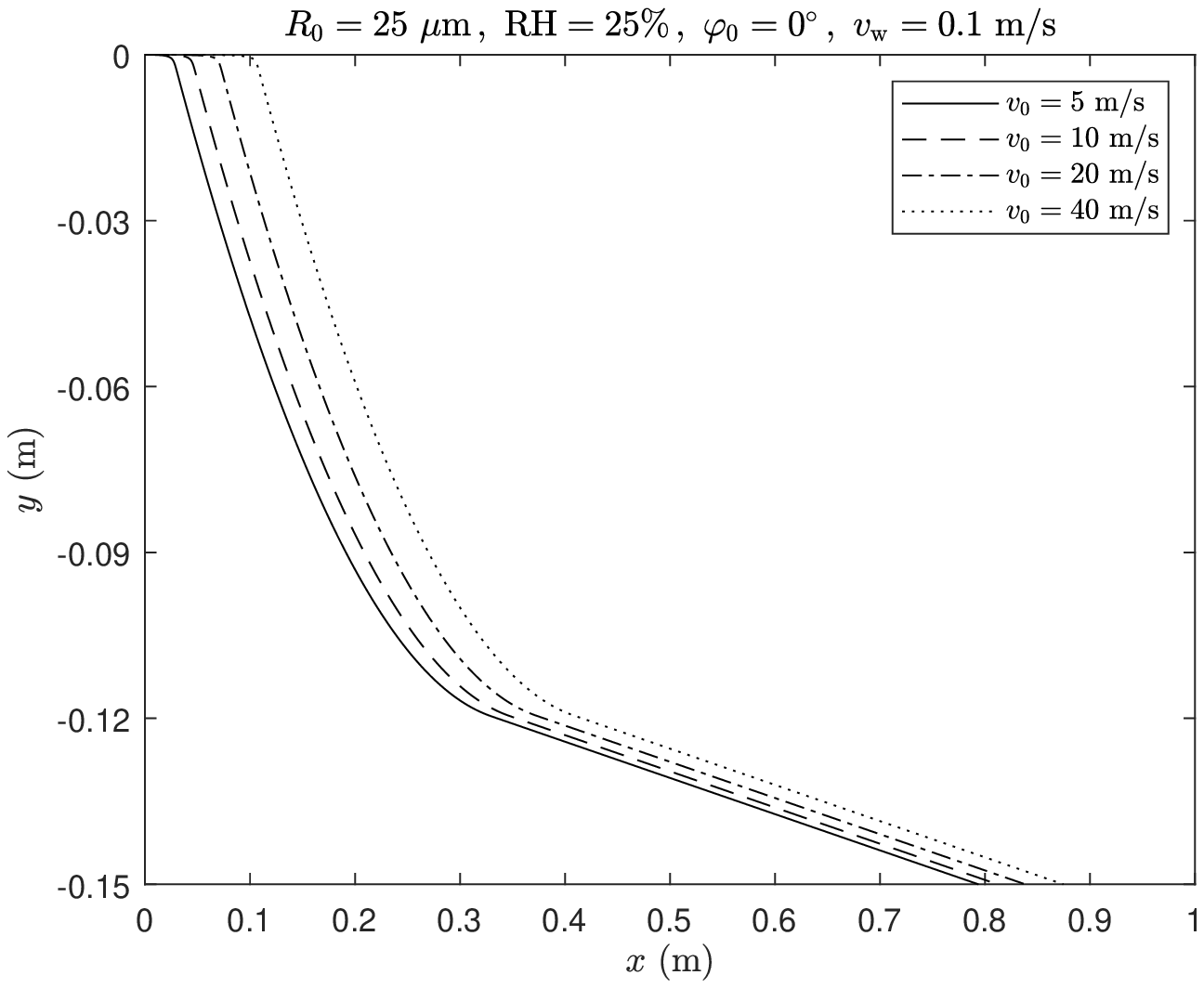}
}
\vskip3mm
\centerline{(a)\hskip80mm (b)}
\vskip3mm
\centering{
\includegraphics[scale=0.45]{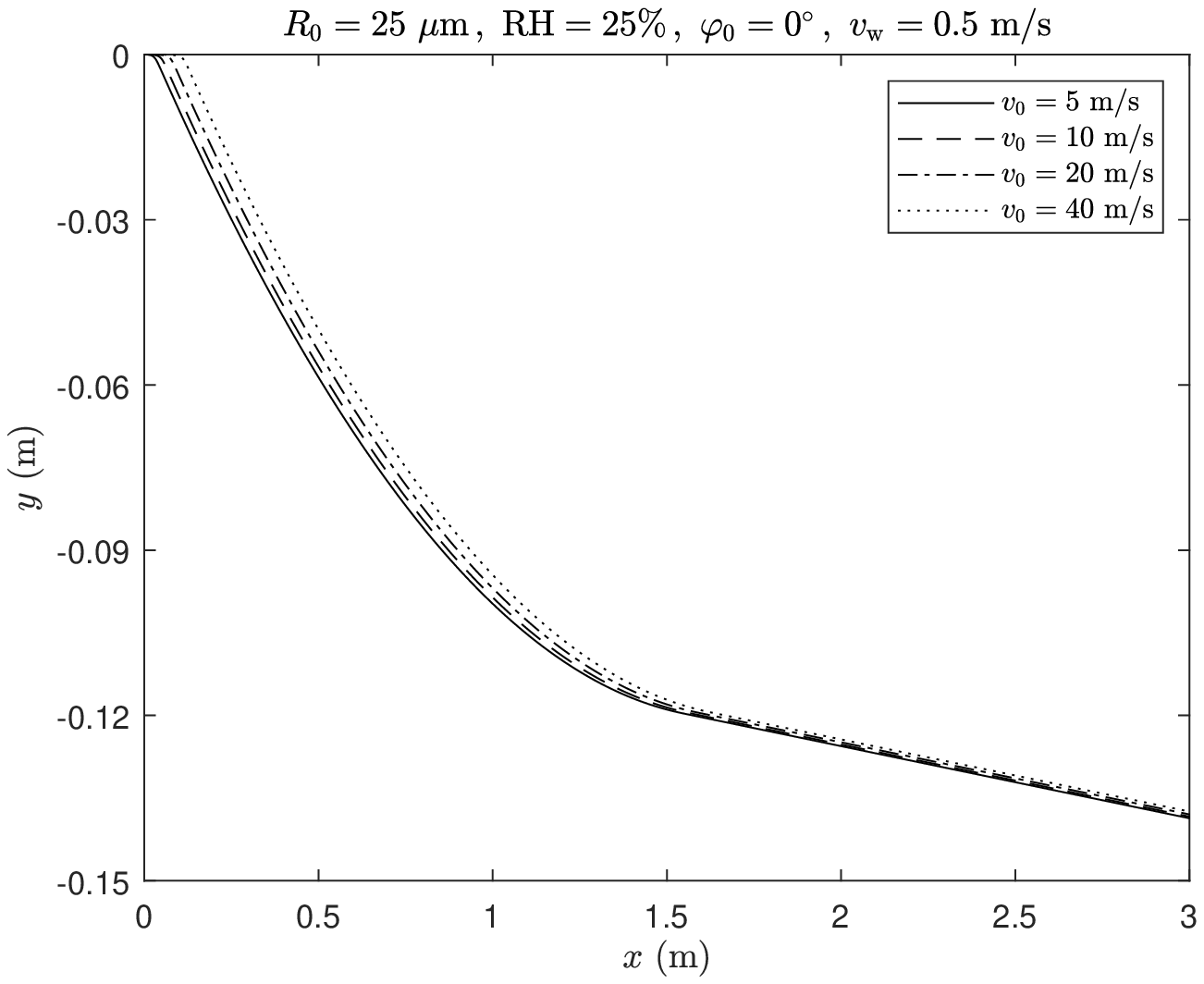}\hskip4mm
\includegraphics[scale=0.45]{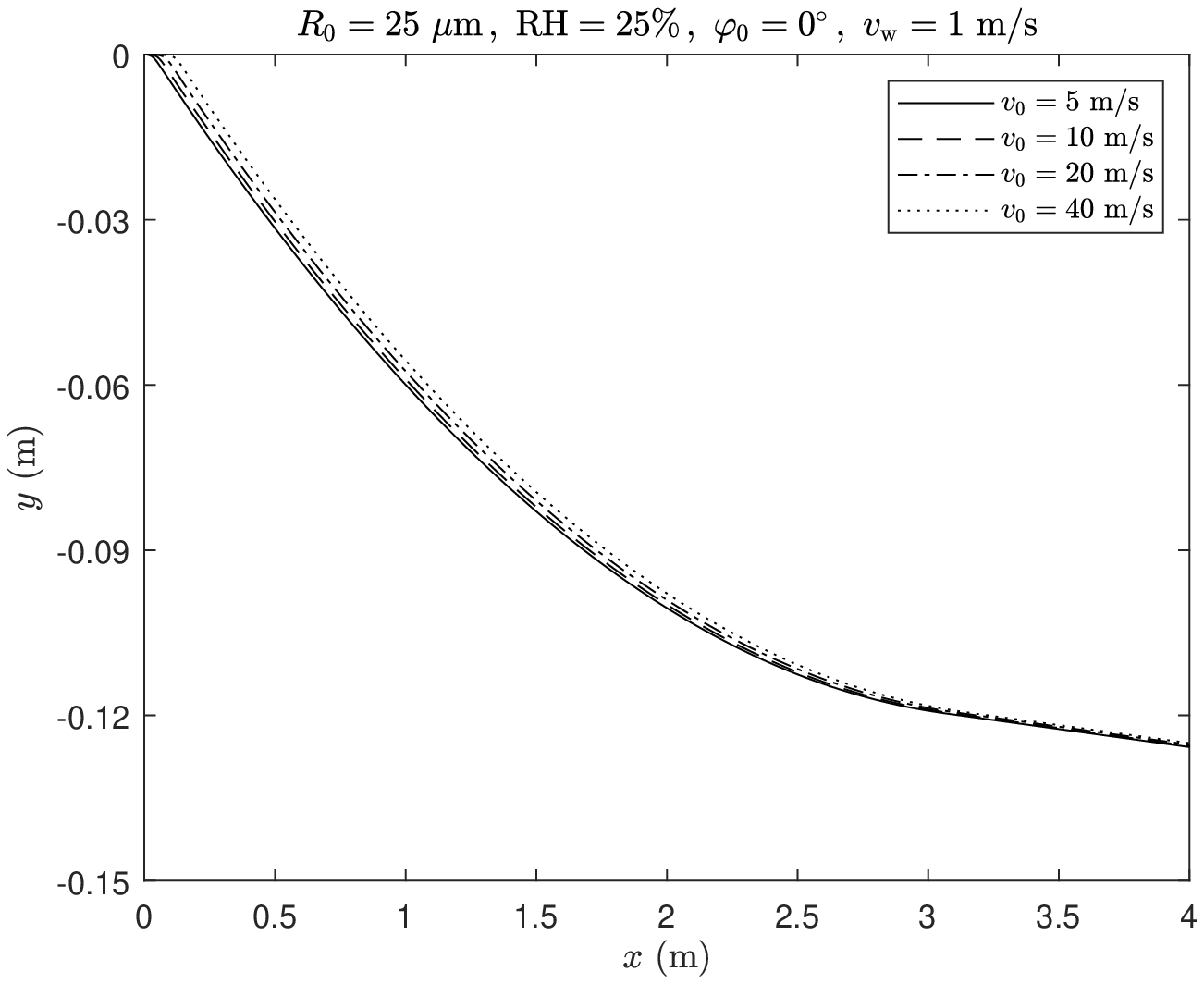}
}
\vskip3mm
\centerline{(c)\hskip80mm (d)}
\vskip2mm
\caption{\label{Fig8} The trajectories of the droplet of initial radius 25 $\mu$m, ejected horizontally  with the velocities $v_0=5$, 10, 20 and 40 m/s.
The wind is absent in part (a), while the wind velocity in parts (b), (c), and (d) is 0.1, 0.5, and 1 m/s, respectively.}
\end{figure}
\section{Wind effects}

The projectile motion considered in previous sections was assumed to take place under quiescent ambient conditions without any external air flow. The effect of external air flow (wind or ventilation) can be readily included in the analysis. For example, if the wind is horizontal, with a constant speed $v_{\rm w}$, the equations of motion are given by (\ref{2.6}), with $v_x$ replaced by $v_x-v_{\rm w}$, and $v$ by $[(v_x-v_{\rm w})^2+v_y^2]^{1/2}$, while the Reynolds number in (\ref{2.3}) becomes ${\rm Re}=2R[(v_x-v_{\rm w})^2+v_y^2]^{1/2}/\nu_{\rm air}$.
Figure \ref{Fig7} shows the the trajectories of the droplet of initial radius $R_0=100\,\mu$m, ejected with four indicated initial velocities in the case of the forward wind with speed 1 m/s and 2 m/s The relative humidity in all cases is assumed to be the same and equal to RH=25\%.
In Fig. \ref{Fig7}a and \ref{Fig7}b, the droplet is ejected horizontally, and in Fig. \ref{Fig7}c and \ref{Fig7}d the ejection angle is $\varphi_0=30^\circ$. If these trajectories are compared with trajectories shown in Fig. \ref{Fig3}b and Fig. \ref{Fig6}a without wind, we can see that the presence of wind greatly affects the shape of trajectory and the maximum horizontal reach of the droplet.
Similarly, Fig. \ref{Fig8} shows the comparison of the trajectories of a droplet with initial radius of only 25 $\mu$m, ejected horizontally in the absence of wind and in the direction
of wind whose velocity is 0.1, 0.5, and 1 m/s. While the maximum horizontal distance in the absence of wind is only about 2.5 to 10.4 cm (depending on the ejection velocity $v_0$),
in the presence of even a mild wind the horizontal reach of the droplet may extend to several meters, while the vertical descent is only about 15 cm. This naturally contributes to the increase of the infection spreading risk, if droplets carry viruses or other pathogens. A recent numerical study of the influence of wind on the COVID-19 airborne transmission has been presented by Feng et al. (2020).

\subsection{Droplet's motion against the wind}

\begin{figure}
\centering{
\includegraphics[scale=0.45]{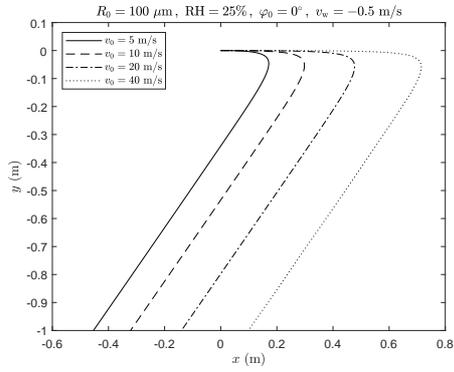}\hskip4mm
\includegraphics[scale=0.45]{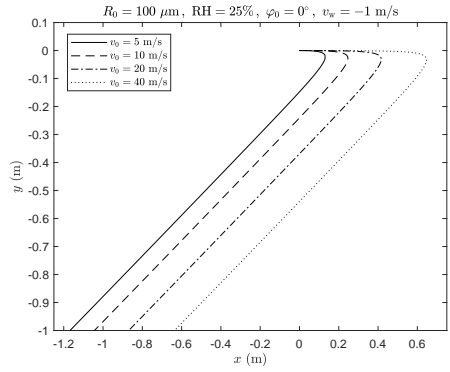}
}
\vskip3mm
\centerline{(a)\hskip80mm (b)}
\vskip3mm
\centering{
\includegraphics[scale=0.45]{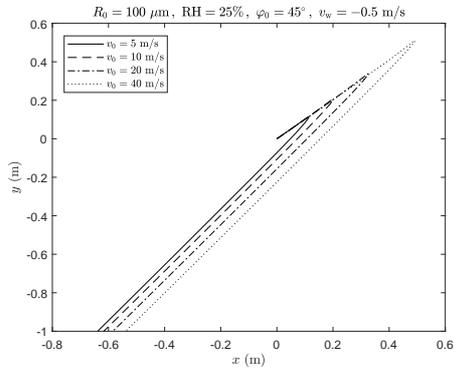}\hskip4mm
\includegraphics[scale=0.45]{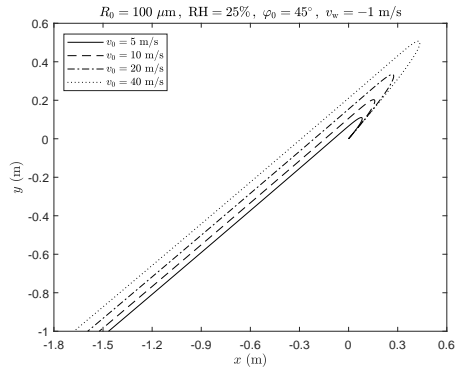}
}
\vskip3mm
\centerline{(c)\hskip80mm (d)}
\vskip2mm
\caption{\label{Fig9} The trajectories of the droplet of initial radius 100 $\mu$m, ejected with the velocities $v_0=5$, 10, 20 and 40 m/s
against the horizontal wind. The direction of $v_0$ is at the angle $\varphi_0=0$ in parts (a) and (b), and $\varphi_0=45^\circ$ in parts (c) and (d).
The wind velocity in parts (a) and (c) is -0.5 m/s, and -1 m/s in parts (b) and (d).}
\end{figure}

Figure \ref{Fig9} shows the trajectories of the droplet of initial radius 100 $\mu$m, as considered in Fig. \ref{Fig7}, in the case of the reversed direction of the wind. Upon initial forward motion, the droplet reverses the direction of its motion and moves backward. Note that, in the case of $\varphi_0=45^\circ$, the reversal of the direction takes place below the line along the direction of the initial speed $v_0$ for the wind speed $v_{\rm w}=-0.5$ m/s, and above that line for the wind speed $v_{\rm w}=-1$ m/s. From all four plots in Fig. \ref{Fig9} it follows that a droplet emitted by a taller person against a wind can easily reach a person standing behind even at distances beyond one or two meters, particularly at stronger wind.

\begin{figure}
\centering{
\includegraphics[scale=0.45]{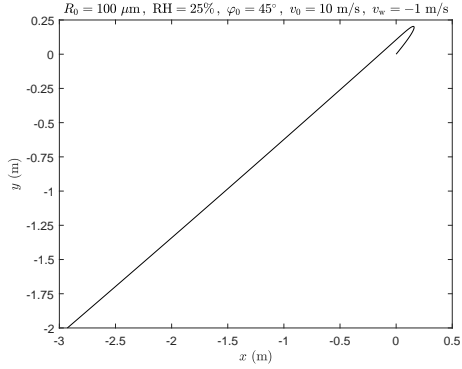}\hskip4mm
\includegraphics[scale=0.45]{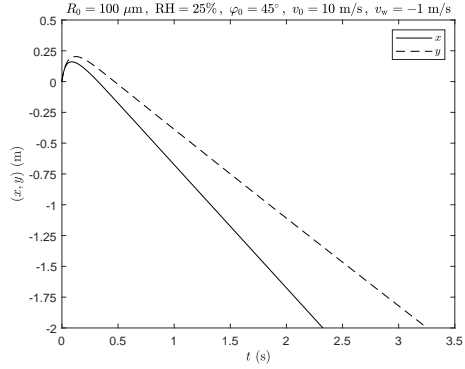}
}
\vskip3mm
\centerline{(a)\hskip80mm (b)}
\vskip3mm
\centering{
\includegraphics[scale=0.45]{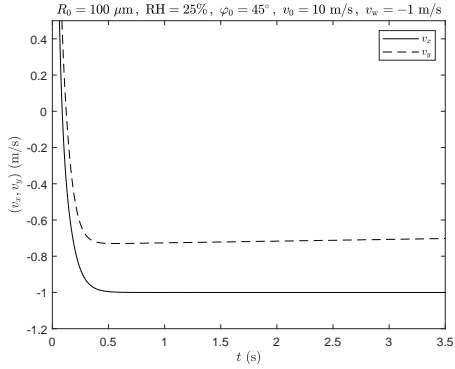}\hskip4mm
\includegraphics[scale=0.45]{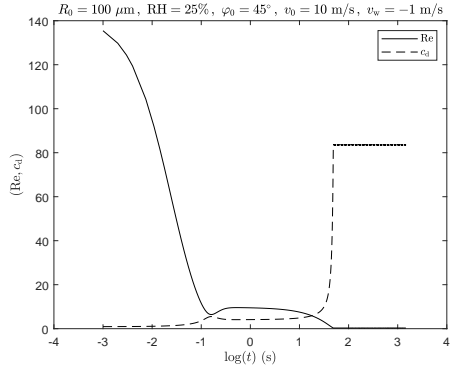}
}
\vskip3mm
\centerline{(c)\hskip80mm (d)}
\vskip2mm
\caption{\label{Fig10} (a) The trajectory $y=y(x)$ of the droplet of initial radius 100 $\mu$m, ejected with the velocities $v_0=10$ m/s at the angle $\varphi_0=45^\circ$
against the horizontal wind $v_{\rm w}=-1$ m/s. The corresponding time variations of (b) the coordinates $x=x(t)$ and $y=y(t)$, and (c)
the velocity components $v_x=v_x(t)$ and $v_y=v_y(t)$. (d) The variations of the Reynolds number and the drag coefficient in the extended time range
(on the logarithmic scale).}
\end{figure}

Figure \ref{Fig10}a shows the trajectory of the droplet of initial radius 100 $\mu$m, ejected at $\varphi_0=45^\circ$ with the initial velocity $v_0=10$ m/s against the wind with velocity $v_{\rm w}=-1$ m/s. Upon quick reversal of its direction of motion, the droplet falls to the ground, 2 meters below the point of ejection, in just 3.2 seconds, far before the droplet would reach its nucleus size of $23\,\mu$m (in 49 seconds). Figure \ref{Fig10}b shows the time variations $x=x(t)$ and $y=y(t)$ during the first 3.5 seconds. The corresponding variations of the velocity components $v_x=v_x(t)$ and $v_y=v_y(t)$ are shown in Fig. \ref{Fig10}c. The horizontal component of the droplet's velocity quickly approaches the value of the wind velocity ($-0.1$ m/s), while the vertical component asymptotically approaches the value of the terminal velocity $v_{\rm t}=0.01$ m/s, albeit in a much longer time (not shown in Fig. \ref{Fig10}c). The slope of the trajectory at such large times would be given by the ratio $-v_{\rm t}/v_{\rm w}$. The variations of the Reynolds number ${\rm Re}$ and the drag coefficient $c_{\rm d}$ with time (on the logarithmic scale, and extended to more than 1,000 seconds) are shown in Fig. \ref{Fig10}d. Only early portion of Fig. \ref{Fig10}d is relevant for the droplet's descent of the order of several or even several tens of meters, because the droplet would descend by 20 m in less than 40 seconds, if not interrupted by the ground or other barrier.

\section{Conclusions}

We have presented an analysis of the projectile motion of a micro-droplet ejected from the mouth at an arbitrary angle and initial velocity, corresponding to soft or loud speaking, coughing or sneezing. It is assumed that a droplet is ejected as an isolated droplet, outside of the puff of exhaled air. The air resistance is represented by a drag force which is a nonlinear function of the droplet's velocity. The evaporation of the  droplet is described by an expression based on the assumption that the rate of decrease of the droplet's external surface area is dependent only on the relative humidity and the ambient temperature. The initial content of the respiratory droplet is taken to be 98 wt\% water, 1 wt\% salt (NaCl), and 1 wt\% protein (mucus). The change of the average density of the droplet caused by evaporation of its water is determined throughout the concentration range, up to the instant when all the water evaporates and the droplet reduces to its nucleus, consisting of salt and dry protein only. The presented density analysis also specifies the change of the droplet's mass during evaporation, which enables the numerical solution of the governing differential equations of the droplet's motion after its emission from the mouth. The trajectories of different-sized droplets ejected at different velocities and under different relative humidities are determined and discussed.
In each considered case the maximum horizontal distance traveled by a droplet is rapidly reached, which is followed by an essentially vertical descent of the droplet in a stagnant air. The droplet's maximum horizontal reach is compared in the case of speaking vs. coughing or sneezing. The time needed for the droplet to evaporate down to its nucleus size is compared with the time for a droplet to reach the ground, after being ejected from a given height. The effect of the rate of droplet's evaporation on the size and the motion of the droplet is quantified.
More rapidly evaporating droplets are lighter and fall to the ground slower, which increases the risk of infection.
The effects of the angle of ejection of the droplet on its maximum horizontal and vertical reach
and the entire trajectory are also evaluated and discussed, which is of importance for the analysis of the transmission of infection in the cases when a person speaks while sitting near a standing person, or vice versa. The negative ejection angles are also of interest in studying the projectile motion of the droplets that have escaped from a moving puff. The wind effect on the droplet's trajectory and its maximum horizontal reach is also studied. For example, while the maximum horizontal distance of the droplet with $25\,\mu$m initial radius in the absence of wind is only about 3 to 10 cm, depending on ejection velocity, in the presence of even a very
mild wind the horizontal reach of the droplet extends to several meters. This naturally contributes to the increase of the risk of infection
spreading, if the droplets carry viruses or other pathogens. If an isolated droplet is ejected against wind, upon its initial forward motion, the droplet reverses the direction of its motion and it moves backward. If a droplet is emitted by a taller person against the wind, it can easily reach a person standing behind even at distances beyond a meter, particularly at stronger wind. The analysis can be readily extended to include other in-plane wind directions. In the case of an out-of-plane wind ($\boldsymbol{v}_0$ and $\boldsymbol{v}_{\rm w}$ being non-coplanar), the projectile motion becomes three-dimensional, which will be discussed elsewhere.
The motion of the droplets that are trapped and carried by exhaled air puff during its forward motion is more involved and not considered in this paper. The motion of small droplets relative to the puff is rather slow, and the model of linear drag may be adopted,
but the main difficulty is the kinetic description of the moving and expanding puff which carries the small droplets, as discussed in a recent review by Balachandar et al. (2020) and the references cited therein.

\section*{References}
\def\ref{\hangindent=7 true mm\hangafter=1\vspace{0pt}\noindent}
\vspace{-1mm} \baselineskip 15.5pt

\ref Anchordoqui, L.A., Chudnovsky, E.M. (2020). A physicist view of COVID-19 airborne infection through convective airflow in indoor spaces. \emph{SciMedicine J.},  2, 68--72.
doi:10.28991/SciMedJ-2020-02-SI-5

\ref Balachandar, S.,  Zaleski, S., Soldati, A.,  G. Ahmadi, G.,  Bourouiba, L. (2020).  Host-to-host airborne transmission as a multiphase flow
problem for science-based social distance guidelines. \emph{Int. J. Multiphase Flow}, 132,  Article 103439. doi:10.1016/j.ijmultiphaseflow.2020.103439

\ref Bourouiba, L. (2020). Turbulent gas clouds and respiratory pathogen emissions: potential implications for reducing transmission
of COVID-19. \emph{JAMA}, 323 (18), 1837--1838.\\
doi:10.1001/jama.2020.4756

\ref Chaudhuri, S.,  Basu, S.,  Kabi, P.,  Unni, V.R., Saha, A. (2020). Modeling the role of respiratory droplets in
Covid-19 type pandemics. \emph{Phys. Fluids}, 32, Article 063309. doi:10.1063/5.0015984

\ref Chen, L.-D. (2020). Effects of ambient temperature and humidity on droplet lifetime - A perspective of exhalation sneeze droplets with COVID-19 virus transmission.
\emph{Int. J. Hyg. Environ. Health}, 229, Article 113568. doi:10.1016/j.ijheh.2020.113568

\ref Cheng, C.H., Chow, C.L., Chow, W.K. (2020). Trajectories of large respiratory droplets in indoor environment: a
simplified approach. \emph{Build. Environ.}, 183, Article 107196.\\
doi:10.1016/j.buildenv.2020.107196

\ref  Das, S.K., Alam, J., Plumari, S., Greco, V. (2020). Transmission of airborne virus through sneezed
and coughed droplets. \emph{Phys. Fluids},  32, Article 097102. doi:10.1063/5.0022859

\ref Duguid, J.P. (1946). The  size  and the  duration  of air-carriage of respiratory droplets  and  droplet-nuclei. \emph{J. Hyg. (Lond.)},  44 (6), 471--4799.
doi:10.1017/s0022172400019288

\ref Feng, Y., Marchal, T.,  Sperry, T., Yi, H. (2020). Influence of wind and relative humidity on the social distancing
effectiveness to prevent COVID-19 airborne transmission: A numerical study, \emph{J. Aerosol Sci.} 147, Article 105585.
doi:10.1016/j.jaerosci.2020.105585

\ref Fischer, H., Polikarpov, I., Craievich, A.F. (2004). Average protein density is a molecular-weight-dependent function. \emph{Protein Science}, 13, 2825--2828.
doi:10.1110/ps.04688204

\ref Giovanni, A., Radulesco, T., Bouchet, G., Mattei, A., Revis, J., Bogdanski, E.,  Michel, J. (2020). Transmission of droplet-conveyed infectious agents such
as SARS-CoV-2 by speech and vocal exercises during speech therapy:
preliminary experiment concerning airflow velocity.  \emph{Eur. Arch. Otorhinolaryngol.}, July 16, 1--6. doi:10.1007/s00405-020-06200-7

\ref  Jakubczyk, D., Kolwas, M.,  Derkachov, G.,  Kolwas, K.,  Zientara, M. (2012). Evaporation of micro-droplets:
the radius-square-law revisited. \emph{Acta Physica Polonica}, 122, 709--716.\\
doi:10.12693/APhysPolA.122.709

\ref Jayaweera, M., Perera, H., Gunawardana, B.,  Manatunge, J. (2020). Transmission of COVID-19 virus by droplets and aerosols: A critical review on the unresolved dichotomy. \emph{Environ. Res.}, 188, Article 109819. doi:10.1016/j.envres.2020.109819

\ref Kayser, R., Jr.,  Bennett, H.S. (1997). Evaporation of a liquid droplet. \emph{J. Res. Natl. Bur. Stand. -- A Phys. Chem.},
81A (2,3), 257--266. doi:10.6028/jres.081A.015

\ref Khan, A.R., Richardson, J.F. (1987). The resistance to motion of a solid sphere in a fluid. \emph{Chem. Engin. Commun.}, 62 (1-6), 135--150.
doi:10.1080/00986448708912056

\ref Kukkonen, J., Vesala, T., Kulmala, M. (1989). The interdependence of evaporation and
settling for airborne freely falling droplets. \emph{J. Aerosol Sci.}, 20, 749--763.
doi:10.1016/0021-8502(89)90087-6

\ref Lapple, C.E. (1951). \emph{Particle Dynamics}. Eng. Res. Lab., E.I. Du Pont De Nemours and Co.,
Wilmington, Delaware.

\ref Li, H., Leong, F.Y., Xu, G., Kang, C.W., Lim, K.H., Tan, B.H., Loo, C.M. (2021). Airborne dispersion of droplets
during coughing: a physical model of viral transmission. \emph{Sci. Rep.}, 11, Article 4617. doi:10.1038/s41598-021-84245-2

\ref Li, X., Shang, Y.,  Yan, Y.,  Yang, L.,  Tu, J (2018). Modelling of evaporation of cough droplets in inhomogeneous humidity
fields using the multi-component Eulerian--Lagrangian approach. \emph{Build. Environ.}, 128, 68--76. doi:10.1016/j.buildenv.2017.11.025

\ref Lieber, C., Melekidis, S., Koch, R., Bauer, H. J. (2021). Insights into the evaporation characteristics of saliva droplets and aerosols: Levitation experiments and numerical modeling. \emph{J. Aerosol Sci.}, 154, Article 105760. doi.org/10.1016/j.jaerosci.2021.105760

\ref Liu, F., Qian, H., Zheng, X., Song, J., Cao, G., Liu, Z. (2019). Evaporation and dispersion of exhaled droplets in stratified environment.
\emph{IOP Conf. Ser. Mater. Sci. Eng.}, 609, Article 042059. doi:10.1088/1757-899X/609/4/042059

\ref Liu, S.,  Novoselac, A. (2014). Transport of airborne particles from an unobstructed cough
jet. \emph{Aerosol Sci. Technol.},  48, 1183--1194. doi:10.1080/02786826.2014.968655

\ref Morawska, L. (2006). Droplet fate in indoor environments, or can we
prevent the spread of infection. \emph{Indoor Air},  16 (5), 335--347. doi:10.1111/j.1600-0668.2006.00432.x

\ref Morawska, L., Johnson, G.R., Ristovski, Z.D., Hargreaves, M., Mengersen, K., Corbett, S.,
Chao, C.Y.H., Li, Y.,  Katoshevski, D. (2009). Size distribution and sites of origin of droplets expelled from the human
respiratory tract during expiratory activities. \emph{J. Aerosol Sci.}, 40 (3), 256--269. doi:10.1016/j.jaerosci.2008.11.002

\ref Mittal, R., Meneveau, C.,  Wu, W. (2020), A mathematical framework for estimating
risk of airborne transmission of COVID-19 with application to face mask  use and  social distancing. \emph{Phys. Fluids}, 32, Article 101903. doi:10.1063/5.0025476

\ref Netz, R.R. (2020). Mechanisms of airborne infection via evaporating and sedimenting
droplets produced by speaking. \emph{J. Phys. Chem. B}, 124, 7093--7101. doi:10.1021/acs.jpcb.0c05229

\ref Papineni, R.S., Rosenthal, F.S. (1997). The size distribution of droplets in the exhaled breath of healthy human subjects. \emph{J. Aerosol Medicine}, 10,
105--116. doi:10.1089/jam.1997.10.105

\ref Sarkar, A., Xu, F.,  Lee, S. (2019). Human saliva and model saliva at bulk to adsorbed phases - similarities and differences. \emph{Adv. Colloid Interface
Sci.}, 273, Article 102034.\\
doi:10.1016/j.cis.2019.102034

\ref Schlichting, H. (1979). \emph{Boundary Layer Theory}, 7th ed. McGraw-Hill, New York.

\ref Stadnytskyi, V.,  Bax, C.E., Bax, A., Anfinrud, P. (2020). The airborne lifetime of small speech droplets and their potential importance in SARS-CoV-2 transmission.
\emph{PNAS}, 117 (22, 11875--11877. doi:10.1073/pnas.2006874117

\ref Su, Y.-Y., Miles, R.E.H., Li, Z.-M., Reid, J.P., Xu, J. (2018). The evaporation kinetics of pure water droplets at
varying drying rates and the use of evaporation rates to infer the gas phase relative humidity. \emph{Phys. Chem. Chem. Phys.}, 20, 23453--23466l.
doi:10.1039/C8CP05250F

\ref Sun, W., Ji, J. (2007). Transport of droplets expelled by coughing in
ventilated rooms. \emph{Indoor Built Environ.},  16 (6), 493--504. doi:10.1177/1420326X07084290

\ref  Tellier, R., Li, Y., Cowling, B.J., Tang, J.W. (2019). Recognition of aerosol transmission of
infectious agents: a commentary.  \emph{BMC Infect. Dis.}, 19,  101--109.
doi:10.1186/s12879-019-3707-y

\ref Vuorinen, V., et al. (2020). Modelling aerosol transport and virus exposure with numerical simulations in relation to SARS-CoV-2 transmission by inhalation indoors.
\emph{Safety  Science}, 130, Article 104866. doi:10.1016/j.ssci.2020.104866

\ref Wang, H., Li, Z., Zhang, X., Zhu, L., Liu, Y., Wang, S. (2020). The motion of respiratory droplets produced
by coughing. \emph{Phys. Fluids}, 32, Article 125102. doi: 10.1063/5.0033849

\ref Wang, Y., Wu, S.H., Yang, Y., Yang, X.N., Song, H., Cao, Z.X., Huang, Y.Q. (2019). Evaporation
and movement of fine droplets in non-uniform temperature and humidity field. \emph{ Build. Environ.}, 150, 75--87. doi:10.1016/j.buildenv.2019.01.003

\ref Wells, W.F. (1934). On air-borne infection study II: droplets and droplet nuclei. \emph{Am. J. Hyg.}, 20, 611--618. doi:10.1093/oxfordjournals.aje.a118097

\ref White, F.M. (2006). \emph{Viscous Fluid Flow}, 3rd ed. McGraw-Hill, New York.

\ref Xie, X., Li, Y., Chwang, A.T.Y., Ho, P.L., Seto, W.H. (2007). How far droplets can move in indoor environments - revisiting the Wells evaporation-falling curve. \emph{Indoor Air} 17, 211--225. doi:10.1111/j.1600-0668.2007.00469

\ref Xie, X.,  Li, Y., Sun, H., Liu, L. (2009). Exhaled droplets due to talking and coughing. \emph{J. R. Soc.
Interface}, 6, S703--S714. doi:10.1098/rsif.2009.0388.focus

\ref Zhu, S., Kato, S., Yang, J.-H. (2006). Study on transport characteristics  of saliva droplets  produced  by coughing  in a calm indoor  environment.
\emph{Build. Environ.}, 41, 1691--1702. doi:10.1016/j.buildenv.2005.06.024

\end{document}